\documentclass[]{interact}
\usepackage{caption}

\usepackage{longtable}
\usepackage[breaklinks=true]{hyperref}
\usepackage{epstopdf}
\usepackage[caption=false]{subfig}
\usepackage[doublespacing]{setspace}
\usepackage{float}
\usepackage{tikz}
\usetikzlibrary{calc,shapes,backgrounds, arrows, positioning, fit}

\usepackage[natbibapa,nodoi]{apacite}
\usepackage{amsmath}
\theoremstyle{plain}

\theoremstyle{definition}

\theoremstyle{remark}

\begin{document}

\articletype{ORIGINAL ARTICLE}

\thispagestyle{empty}
	
	 \newgeometry{left=2cm,right = 2cm, bottom = 2 cm}

\title{Missing Data in Discrete Time State-Space Modeling of Ecological Momentary Assessment Data: A Monte-Carlo Study of Imputation Methods}

\author{
Lindley R. Slipetz \\
\affil{University of Virginia, Charlottesville, USA}
Ami Falk\\
Teague R. Henry\\
\affil{University of Virginia, Charlottesville, USA}
}

\maketitle

\begin{abstract}
When using ecological momentary assessment data (EMA), missing data is pervasive as participant attrition is a common issue. Thus, any EMA study must have a missing data plan. In this paper, we discuss missingness in time series analysis and the appropriate way to handle missing data when the data is modeled as an idiographic discrete time continuous measure state-space model. We found that Missing Completely At Random, Missing At Random, and Time-dependent Missing At Random data have less bias and variability than Autoregressive Time-dependent Missing At Random and Missing Not At Random. The Kalman filter excelled at handling missing data under most conditions. Contrary to the literature, we found that, using a variety of methods, multiple imputation struggled to recover the parameters. 
\end{abstract}

\begin{keywords}
missing data; time series; ecological momentary assessment; state-space model
\end{keywords}

\newpage

\section{Introduction}

\textbf{This article has been accepted for publication in Multivariate Behavioral Research, published by Taylor \& Francis.}

Missing data is ubiquitous in psychological data, be it cross-sectional or across time. This is problematic as many statistical methods are not directly applicable to data sets with missing data, and those that are face problems with statistical power, validity, and accuracy of parameter estimates when missing data is handled inappropriately \citep{el-masri_missing_2005}. There is often confusion regarding determining the pattern of missingness from the original patterns in \citet{rubin_inference_1976}, further bolstering the above problems in that if certain types of missingness are ignored, they will result in severely biased parameter estimates, which in turn can result in spurious findings. Even when correctly determining the mechanism of missingness, the volume of missing data imputation methods available may leave psychological researchers puzzled about which method is most appropriate for their data.

While missing data occurs and creates problems in the cross-sectional case, this is amplified in the case of intensive longitudinal or time series data, particularly ecological momentary assessment (EMA) data. EMA data is time series data that is collected at multiple time points in the participant's natural environment, usually multiple times a day over several weeks \citep{shiffman_ecological_2008}. While EMA is popular in clinical populations, it is capable of measuring any phenomenon that is thought to change over time such that daily patterns may differ from weekly patterns or within day patterns differ from between day patterns. For example, we might look for trends in cigarette smoking, mood, or social activities.  In the case of cigarette smoking, you may see differences within a day (e.g, smoking less in the morning) or differences throughout the week (e.g., smoking more on the weekend than on weekdays). Similar patterns may hold for the case of social activities as well. EMA studies are said to be ecologically valid in that data is collected in natural environments rather than in a laboratory setting \citep{shiffman_ecological_2008}. It also avoids issues with participants having to remember thoughts, feelings, and behaviors as they would if they were questioned in a laboratory setting because EMA can ask about the events in real-time. A major challenge of EMA data collection is participant attrition. To adequately model the data, we need participants to faithfully respond to the measures for a large number of time points. As participants do not typically fulfill this ideal, the EMA researcher must have a response to the problem of missingness. 

There are a variety of ways of modeling EMA data, from observed variable time series models, to the use of mixed effects models. One general framework for analyzing the temporal dynamics of EMA data is state-space modeling. A state-space model is one where the observed variables, in this case the individual items collected at each timepoint in a EMA study, are considered to be measurements of an underlying latent process (in state-space parlance, these latent processes are the states). In this framework, the temporal dynamics of the EMA data are fully captured in the temporal relations between these latent states. 

Here we use an idiographic discrete time continuous measure state-space model, represented via the following equations:

\begin{gather}
\mathbf{x}_{t+1} = \mathbf{A}\mathbf{x}_{t} + \mathbf{v}_{t}\\
\mathbf{v}_{t} \sim N_{p}(\mathbf{0}_{p}, \mathbf{Q})\\
\mathbf{z}_{t} = \mathbf{H}\mathbf{x}_{t} + \mathbf{w}_{t}\\
\mathbf{w}_{t} \sim N_{p}(\mathbf{0}_{p}, \mathbf{R})
\end{gather}

where there are $p$ states, $T$ time points, and $l$ indicators, $\mathbf{X}_{t}$ is the $p \times 1$ vector of latent states at time $t$, $\mathbf{A}$ is the $p \times p$ matrix of transition coefficients for the latent states which contains autoregressive effects (i.e., how the variable relates to itself, dynamically) on the diagonal and cross-lagged effects (i.e., how the variable relates to another variable, dynamically) in the other cells, $v_{t}$ is the multivariate normally distributed disturbance term for the latent states with mean 0 and $p \times p$ covariance matrix $\mathbf{Q}$, $\mathbf{Z}_{t}$ is the $l \times 1$ vector of measurements at time $t$, $\mathbf{H}$ is the $l \times p$ matrix that maps states to measurements, $w_{t}$ is the multivariate normally distributed measurement error for the measured variables with mean 0 and $l \times l$ covariance matrix $\mathbf{R}_{t}$.

This idiographic model was chosen due to its simplicity, acting as a first step in exploring the role of missing data in EMA analysis via state-space models. This discrete time model assumes equal intervals of data collection, an idealization of EMA data collecting processes. In addition, we assume that the measured variables are continuous, another simplifying assumption. These idealizations serve to help assess the impact of missing data in EMA studies at a basic level.

In addition, the above state space equations can be modified to accommodate an explicit time variable. The term, $\mathbf{G}\mathbf{u}_t$, can be added to equation 1 in which $\mathbf{u}_t$ is a vector of observed exogenous predictors (e.g., an external event, age, or the day of the week) and $\mathbf{G}$ is a coefficient matrix that maps the effects of the exogenous predictors (e.g., the effect of a scheduled intervention) onto the states. Thus time can be coded as an exogenous variable and included in $\mathbf{u}_t$, allowing analysts to include developmental trajectories or seasonal effects with ease.

As an example of state-space modeling at work, \citet{mckee_emotion_2020} presented a state-space model in the psychological context, a feedback-control model of homeostasis that serves to develop an index of an 18 time point time series of 11 affect items for emotion regulation in a high-risk adolescent sample (i.e., ages 13-14). While this is an example of a continuous time model, it demonstrates the capabilities of state-space modeling generally. In the model, the affective state, $\mathbf{x}$ (measured by the above observed affect measures,$y_t$ in the original paper, here $\mathbf{z}_t$ in Eq. 3) and an autoregressive regulation parameter, ($\lambda$ in the original paper, here this parameter would be on the diagonal of the $\mathbf{A}$ matrix), that acts as index of emotion regulation (i.e. how quickly affect returns to equilibrium), and the unmeasured influences on the individuals (i.e., the standard deviation of exogenous disturbances, $\mathbf{u}$) were included. Using a cross-sectional, between-person analysis of the total counts of nicotine, alcohol, and cannabis use with the idiographically calculated autoregressive regulation parameter and means/standard deviations of the sum scores of the affect measures as predictors, they found significant inverse relationship between the affect regulation parameter and substance use, suggesting that participants who take longer to return to their affective baseline tend to have increased substance use. These results serve as an example of state-space modeling’s ability to track affective regulation over time, specifically, or trends in psychological phenomena over time, generally.

The purpose of this paper is to understand the impacts of different types of missing data on the analysis of EMA-like data and offer guidelines for the use of missing data imputation methods in those cases.  First, we review missing data mechanisms and discuss how they can occur in EMA data. Second, we perform a Monte-Carlo simulation study of the impacts of different missingness mechanisms on the statistical modeling of EMA-like synthetic data, comparing the ability of several missing data imputation techniques in the time series case. We conclude by providing guidance to the psychological researcher regarding appropriate missing data approaches.

\subsection{Mechanisms of Missingness}

Missing data mechanisms are traditionally divided into three categories, Missing Completely At Random (MCAR), Missing At Random (MAR), and Missing Not At Random (MNAR), which first appear in \citet{rubin_inference_1976}. These three broad categorizations describe how the probability of missingness is or is not related to the variables in the dataset. To define our notation, let $Y$ be a variable that has missing values, with $Y_{mis}$ and $Y_{obs}$ referring to the values of $Y$ for the missing and observed cases respectively, and let $X$ be a different observed variable.

\begin{figure}[H]
\centering
\includegraphics[width=.9\textwidth]{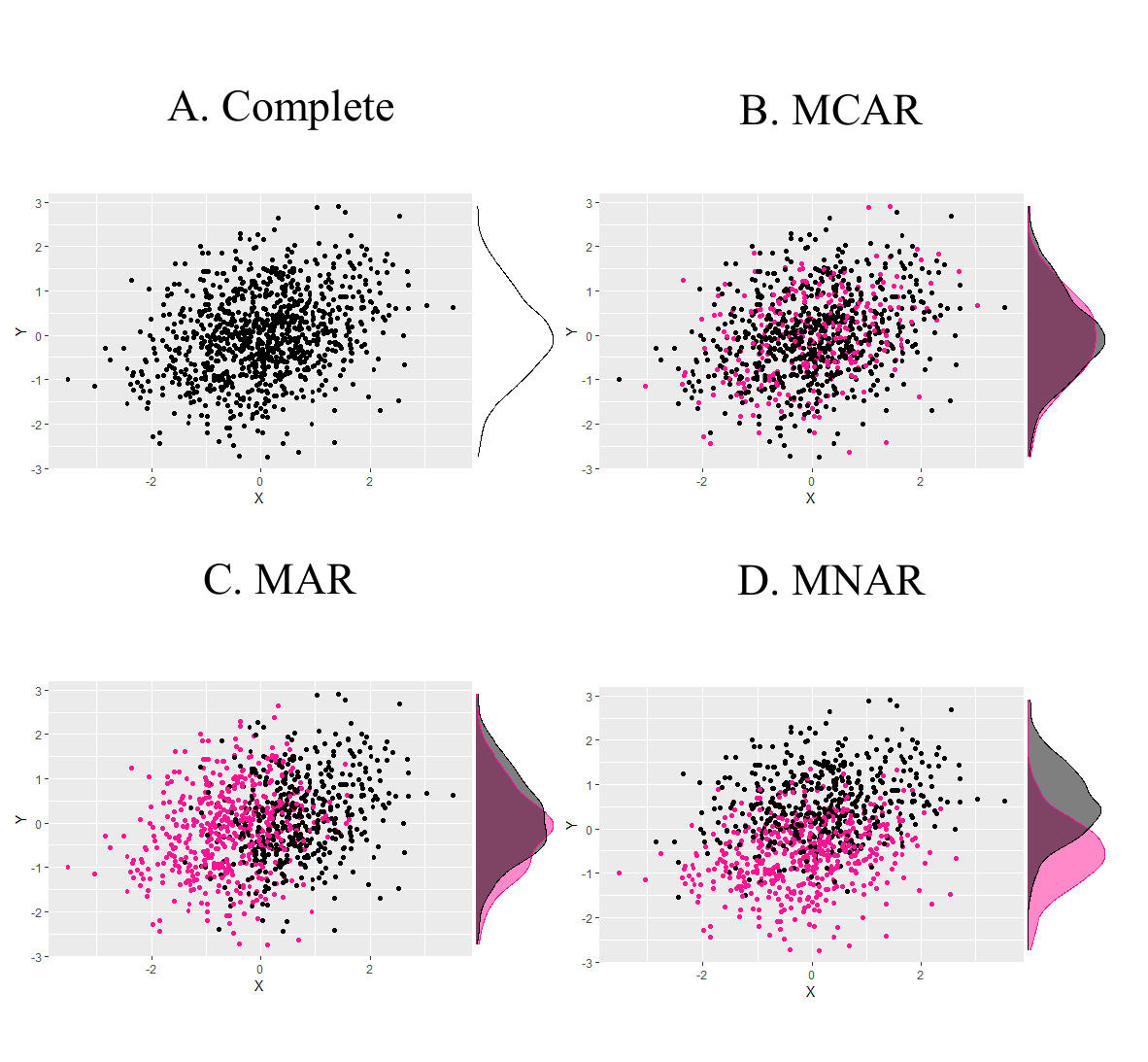}
\caption{The above graphs show variables correlated at .3 with A. complete data, B. MCAR-deleted data, C. MAR-deleted data, and D. MNAR-deleted data. The black dots represent the data that is present in the sample and the pink dots represent data that is missing. Variable $Y$ is the variable with missing data (in B-D) and variable $X$ is a variable with complete data. Note that in Panel B, the marginal distributions of the MCAR missing data are the same as the marginal distributions of the non-missing data, while in Panel C and D, the marginal distributions of the non-missing data are considerably different from the original complete data marginal distribution.}
\label{fig:missing}
\end{figure}

MCAR is the simplest missingness mechanism, and describes the case where the probability of missing data is completely unrelated to the values of data: $P(Missing | Y_{mis}, X) = P(Missing)$. MCAR, while being the simplest missingness mechanism mathematically, is also the most unrealistic missingness mechanism to find in real data, and tends to only occur in planned missingness designs \citep{little_joys_2014}. For a visualization of MCAR data, see Panel B of Fig \ref{fig:missing}. MAR and MNAR describe missingness mechanisms that are dependent on values of the data itself. MAR describes the case where the probability of missing data is dependent on the values of \textit{observed data}: $P(Missing | Y_{mis}, X) = P(Missing|X)$.A visualization of MAR data can be found in Panel C of Fig \ref{fig:missing}. MNAR describes the case where the probability of missing data is dependent on the values of the \textit{missing data itself}: $P(Missing | Y_{mis}, X) \neq P(Missing|X)$. Unlike with MAR, where there is observed data that enables the direct modeling of the missingness mechanism, MNAR mechanisms are considerably more difficult to account for as the information necessary to model the missingness process is itself missing.

There are various methods for determining whether missingness follows the MCAR or MAR mechanisms, though this is not possible for MNAR missingness without access to information about the missing data. To assess MCAR missingness, the test presented in \citet{little_test_1988} is suggested. For MAR missingness, there is no one way that is broadly accepted across the literature. One common method is to model the missingness directly, for example, by coding the missingness of a particular variable and regressing this missingness indicator on potential predictors (using a logistic model).  There are several R packages available for exploring MAR missingness, including via testing in the finalfit R package \citep{finalfit} or visually via the naniar \citep{naniar}, ggplot2 \citep{ggplot2} or visdat \citep{visdat} R packages.

While MCAR, MAR and MNAR mechanisms apply to time series data, their construction doesn't explicitly include time as a possible component of the missingness mechanism. Here, we propose two additional categorizations of missingness mechanisms that are tailored to time series and longitudinal data, time-dependent missing at random (TMAR) and autoregressive time-dependent missing at random (ATMAR).

TMAR occurs when the missingness depends only on the time variable in the model. Similar to MAR missingness, it can be expressed by $P(Missing|Y_{obs,{t}}, Y_{miss,{t}}, X_t, t) = P(Missing|t)$. We can define this mechanism for any timescale: response rates might be lower later during the day, during a weekend, or towards the end of the study period due to participant exhaustion.  These scenarios occur commonly in EMA studies when data is collected very frequently within a day. Importantly, TMAR is a subcategory of MAR, as time is an observed variable. Therefore, missing data models can condition on an appropriately coded time variable and account for TMAR.

Our final category of missing data mechanism, \textit{autoregressive TMAR} or ATMAR, occurs when the missingness on a variable at time $t$ is dependent upon the previous values of the variable: $P(Missing_t | Y_{t}, Y_{t-1}, X_t) = P(Missing_t | Y_{t}, Y_{t-1})$ or $P(Missing_t | Y_{t-1})$ This category is a mix of both MAR and MNAR, as previous timepoints of a given variable may be missing or not missing. For example, consider a study that uses EMA to collect depression severity information. Depression severity tends to be consistent across smaller timescales, meaning that a high value of depression at time $t$ predicts a similar value of depression at time $t+1$. If missingness in the depression variable is dependent on the severity of depression, then the missingness mechanism can be categorized as both ATMAR and MNAR, as information about the value of depression at time $t+1$ is provided by the value of depression at time $t$. Importantly, it is unlikely that a missingness mechanism at $t+1$ can be wholly characterized by the values of the variable at $t$, and it more likely that the ATMAR scenario arises when the mechanism is truly MNAR. However, unlike in a MNAR scenario with cross-sectional data, where there is no information about the values of the missing variables, in a time-series scenario previous values of the variable might be observed and allow analysts to use them to model missingness. The ATMAR mechanism straddles the lines between MAR and MNAR mechanisms.

In the context of EMA, these missing data mechanism occur with varying frequency. As with cross-sectional MCAR, time series MCAR is relatively rare as the criterion is difficult to meet. MAR and MNAR will occur in time series at the same frequencies as their cross-sectional counterparts, in both cases being relatively common. TMAR is the most commonly occurring missing data mechanism in EMA as it depends on the data collection schedule (and it is rare to have a perfect data collection schedule). Finally, ATMAR would occur when there is an autoregressive relation for a given variable, and that variable is afflicted by MNAR missingness.

While there are commonalities between missingness that occurs in cross-sectional data and time series data, time-dependent missingness is unique to time series analysis. Because of this, methods of missing data imputation differ between cross-sectional and time series data. This will be considered in the next section.

\subsection{Missing data methods for time series}
There are a number of methods proposed for handling missing data. The least sophisticated of these is known as listwise deletion, in which all variable values for a given participant at time $t$ are deleted if any one of those variables has missing data. Additionally, mean, mode, or median imputation is often used to replace the missing value with the relevant summary statistic. 
Generally, these methods are inappropriate for any form of missingness other than MCAR due to differences between the observed and complete distributions, but they are particularly unsuitable for time series data because they ignore the time-dependency of the data. Additionally, deletion would fail to preserve the equal intervals required by discrete time models, creating further problems. Here, we will consider multiple imputation methods, which use information from the observed other variables in the model to inform the generation of the missing data imputation values, multiple imputation by chained equations \citep[MICE: ][]{ji_handling_2018}. We will also consider single imputation methods which take into account the temporal dependency in EMA data: the Kalman filter and a variety of methods using the Expectation Maximization algorithm \citep[EM: ][]{dempster_maximum_1977}, chosen due to their citation popularity and the availability of $R$ packages.

Previous work has been done on missingness in repeated measures designs. \citet{diggle1994} present a combined analysis/dropout model for repeated measures data, which, while certainly conceptually applicable to the missing data issues we are studying, is not directly applicable to state-space modeling. Such an application would require a substantial modification of the state-space model and estimating routine. Similarly, the random-coefficient pattern-mixture models proposed in \citet{little1995} have a similar issue, requiring the integration of a missing data model with the state-space model. This sort of bespoke modeling of missing data in this context is of interest, but beyond the scope of our evaluation of existing, readily implementable missing data approaches. As they are not implemented in R, we did not include them in the simulation study.

\subsection{MICE}

 MICE \citep{ji_handling_2018} is a multiple imputation method in which missing data values are drawn through variable-by-variable iterations over conditional densities with Markov Chain Monte Carlo (MCMC) techniques. It is an extraordinarily popular and recommended multiple imputation method and hence, here, we assess its applicability within the context of state-space modeling. Mathematically, consider $\mathbf{Y}$, to be a matrix of dependent variables for all individuals and $\mathbf{X}$ to be a matrix of the covariates for all individuals. Then, $\mathbf{Y}_{o}$ and $\mathbf{X}_{o}$ are the observed variables and $\mathbf{Y}_{m}$ and $\mathbf{X}_{m}$ are the missing variables. With $\theta$ as the vector of unknown parameters, the $i$th iteration of the chained equation is a Gibbs sampler, a MCMC algorithm for sampling from a multivariate distribution, that iteratively draws from the distributions
\begin{align}
\mathbf{\theta}_{i} &\sim P(\mathbf{\theta}|\mathbf{Y}_{o}, \mathbf{X}_{i-1})\\
\mathbf{Y}_{m(i)} &\sim P(\mathbf{Y}|\mathbf{Y}_{o}, \mathbf{X}_{i-1}, \mathbf{\theta}_{i})\\
\mathbf{X}_{m(i)} &\sim P(\mathbf{X}|\mathbf{X}_{o}, \mathbf{Y}_{i}, \mathbf{\theta}_{i})
\end{align}

Here, the draws for the missing variables, $Y_{m(i)}$ and $X_{m(i)}$ are conditional on the non-missing dependent and independent variables, respectively, and the complete imputed data from the independent and dependent variables, respectively, and the draws from the parameter distribution. It is assumed that the relations among the variables follow a general or generalized linear model, though there is flexibility depending on the distributional characteristics of the data. Another benefit of MICE is that, unlike many other imputation methods, it can handle a mix of variable types (e.g., continuous, ordinal, nominal, etc.).

The default imputation method in MICE for numerical variables is \textit{predictive mean matching} \citep[PMM:][]{van_buuren_mice_2011, rubin_multiple_1986, little_test_1988}
. To summarize the relevant details of this approach, PMM begins by first fitting Bayesian regression models predicting the target variable $Y$ by the covariates $X$ (in this application, these are linear regression models) to obtain the posterior distributions of regression parameters. Then, for each given missing value of $Y$, $Y_j[mis]$, pairwise distances are calculated as $|X_i[obs]\hat{\beta} - X_j[mis]\dot{\beta}|$. From there, candidate observed values of $Y$ are selected from cases with the minimum calculated distance. The missing value $Y_j[mis]$ is then imputed by randomly sampling one of those observed values (with probability proportional to the distance metric). Note that this method imputes missing values with values drawn from observed data, rather than sampling from some theoretical distribution. In the context of the state-space models studied here, this implies that MICE, at least used in a typical manner, does not take into account any latent state information (which contrasts this approach with the Kalman filter described below). Also note that, by default, this approach does not account for temporal dependency when applied to timeseries data, as missing values are predicted by other variables at the same timepoint. However, by including lagged versions of variables as additional predictors, the PMM approach can include temporal information. We explore both the standard PMM, MICE-def,  and the use of lagged predictors, MICE-t, in our simulation study.

In addition to the predictive mean matching method, there is also the Bayesian linear regression method, also called the normal method. This approach performs multiple imputation of missing observations which are normally distributed, using a linear regression model with other variables included as predictors (in our case, the non-missing measurements). The main difference between the normal method and the PMM is that while PMM imputes using observed values, the normal method imputes via draws from a Gaussian distribution, thus potentially resulting in values that do not occur in the data. As with our use of PMM, in our simulation study we evaluate conditions where we use the normal method with only the variables from a given timepoint, MICE-norm, and where we use both variables from the current timepoint and the lagged values, MICE-tn.

Simulation studies on the procedure \citep{ji_handling_2018} have shown recovery of point estimates with less bias than listwise deletion. MICE results in multiple imputed datasets, requiring the aggregation of results across datasets during analysis. It has also seen success with binary responses \citep{hardt_multiple_2013, zaninotto_missing_2017}, in models with interactions \citep{zaninotto_missing_2017}, and when imputing missing scale items not the individual-level, but at the level of scale summaries \citep{plumpton_multiple_2016}.

An alternative multiple imputation method not considered here is the the \{Amelia\} R package \citep{honaker_amelia_2011}, which is able to fit polynomial trends to timeseries data. This would improve imputation in situations where there are clear polynomial trends, such as panel longitudinal studies, where the trajectory over time is often of greatest interest. However, Amelia II is limited to deterministic time trends, which are separate from the autoregressive effects of interest here. Hence, we do not consider Amelia here.

As MICE is a multiple imputation method, a method that imputes $i$ many datasets, it requires some way of compiling and comparing the multiple imputed data sets. The standard way appears in \citet{rubin_multiple_1996} and is summarized here (see Supplemental Material for details). The actual posterior distribution defined by $\theta$ is the average of the posterior predictive distributions of the missing data, given by the observed data. It follows that the posterior mean of the distribution defined by $\theta$ is the average of the posterior means of the multiply imputed data sets and the variance of the posterior distribution defined by $\theta$ is the average of the variances of the multiply imputed data sets plus the variance of the multiply imputed data sets. In this way, $i$ many data sets which have been multiply imputed can be analyzed as one data set.

\subsection{Expectation Maximization Based Filtering }

\citet{junger_imputation_2015} propose an approach that modifies the expectation-maximization algorithm of \citet{dempster_maximum_1977} to account for the dependency between timepoints in time series data. The EM algorithm has found broad and popular application across quantitative psychology and, thus, we believe it shows enough promise within the current context to supply a promising condition for our simulation study. 

The approach can be summarized as follows: In the expectation step, missing measurements at time $t$, $\mathbf{\tilde{z}}_t$ are imputed as conditional means using information from non-missing measurements at time $t$, $\mathbf{z}_t$, and estimates of the mean and covariances of $\mathbf{\tilde{z}}_t$ and $\mathbf{z}_t$ using the following expression

\begin{equation}
    \mathbb{E}[\mathbf{\tilde{z}}_t | \mathbf{z}_t, \tilde{\mu}_t, \mu_t, \mathbf{\tilde{\Sigma}}, \mathbf{\Sigma}] = \tilde{\mu} + \mathbf{\tilde{\Sigma}}\mathbf{\Sigma}^{-1}(\mathbf{z}_t - \mu_t)
\end{equation}
where $\tilde{\mu}_t$ is the estimated mean of the missing data at time $t$, $\mu_t$ is the estimated mean of the observed measurements at time $t$, $\mathbf{\tilde{\Sigma}}$ is the estimated covariance between the missing variables and non-missing variables at time $t$, and $\mathbf{\Sigma}$ is the estimated covariance matrix for the non-missing variables at time $t$.

The aim of the maximization step is then to take the imputed time series from the expectation step and estimate the means and covariances of the missing and non-missing measurements. \citet{junger_imputation_2015} provide three different estimation models that are tested here. First, an autoregressive integrated moving average (ARIMA) model can be used by finding the d-th order difference and using the mean estimate of the one-step ahead prediction. Second, a natural cubic spline with curve, $g$, can find an initial mean as $g(x_{t})$ gives an estimate. Finally, \citet{junger_imputation_2015} propose the use of a generalized additive model \citep[GAM:][]{trevor_hastie_generalized_1986}, which are flexible regression type models that allow for non-linear relations to be automatically fit. 

The expectation and maximization steps are performed until convergence, at which time the final timeseries, consisting of non-missing measurements and imputed conditional expected values of the missing measurements, is used in the estimation of the analysis state-space model. We note here that unlike with the unmodified MICE approaches, this EM-based imputation method does take into account the temporal dependency of observations via the use of some filtering method in the maximization step. 

Finally, we must note that the EM based filtering approach developed by \citet{junger_imputation_2015} is a single imputation approach. Most single imputation approaches to missing data lead to bias in subsequent analysis due to both underestimated variance in the imputed missing data, as well as bias in the estimated conditional means \citep{von_hippel_biases_2004}, which can occur in both nomothetic and idiographic models. Furthermore, choice of imputation model can effect the introduction of bias. However, the EM approach to single imputation is a method that leads asymptotically unbiased estimates of the conditional means \citep{von_hippel_biases_2004} but, as with other single imputation approaches, the variance of the missing data is still underestimated.

\subsection{Kalman Filtering} In the current work, we treat 
Kalman filtering as our baseline method for handling multivariate missing time series data modeled as a state-space model as it is a general-use estimation technique for state-space modeling that is readily available via R packages. At the conceptual level, Kalman filtering is a method for estimating the expected values of the latent states at a given timepoint from all observed measurements up to and including that timepoint (i.e. $\mathbb{E}[x_t|z_{1:t}]$). Kalman filtering is part of the core estimation loop for discrete time state-space models, as it not only estimates the expected values of the states, but also permits an easy calculation of the log-likelihood of the overall model. While it is not a missing data imputation method in the traditional sense in that it does not impute values for the observed variables, it can be used in the presence of missing data without requiring any imputation. To elaborate,  Kalman filtering, at time $t$, recursively estimates latent states, $x_{t}$, and a covariance matrix of states, $P_{t}$. 

The first phase of Kalman filtering provides an estimate of the expected state values and state covariance matrices using only the previous timepoints estimate of the states and covariance matrix 

\begin{gather}
\bar{\mathbf{P}}_{t+1} = \mathbf{A}\mathbf{P}_{t}\mathbf{A}' + \mathbf{Q}\\ 
\bar{\mathbf{x}}_{t+1} = \mathbf{A}\mathbf{x}_{t}.
\end{gather}

Note that Eq 9 and 10 do not use any of the observed measurements at time $t$. When there are observed measurements at time $t$, Kalman filtering updates the previous estimates using measurements as follows:

\begin{gather}
\mathbf{P}_{t+1} = [(\bar{\mathbf{P}}_{t+1})^{-1} + \mathbf{H}'\mathbf{R}^{-1}_{t+1}\mathbf{H}]^{-1}\\
\mathbf{x}_{t+1} = \bar{\mathbf{x}}_{t+1} + \mathbf{P}_{t+1}\mathbf{H}'\mathbf{R}^{-1}(\mathbf{z}_{t+1} - \mathbf{H}_{t+1}\bar{\mathbf{x}}_{t+1})
\end{gather}

When $\mathbf{z}_{t+1}$ is missing, an estimate of the expected state value and covariance matrix can still be calculated using Eqs 9 and 10. Hence, we can still use the Kalman filter on timeseries with missing data. In the language of missing data methods, we characterize Kalman filtering as a single pass imputation method at the level of the states, rather than imputing the observed measurements. Notably, Kalman filtering is the only method we evaluated that takes advantage of the use of latent states in the model structure. The variants of MICE and the EM filtering approaches we evaluated here all operate on the level of the observed measurements, posing imputation models at that level. Furthermore, Kalman filtering explicitly takes into account the temporal dependence between timepoints (at the level of the states). Finally, one important drawback to the Kalman filter approach to missing data is that it is only applicable to state-space models, while other missing data approaches evaluated here are general use methods.

The preceding missing data imputation methods provide both single and multiple imputation approaches to time series data, taking into account the time dependencies characteristic of such data. In the context of cross-sectional data, multiple imputation tends to be recommended over single imputation \citep{sinharay_use_2001}. In the case of single imputation, subsequent analyses on the imputed data will be biased by the underestimation of variation in the imputed missing data, while, for multiple imputation, as many data sets are estimated, the uncertainty in the missing data imputation can be accounted for \citep{donders_review_2006}. In the proceeding section, we propose a simulation to examine if this recommendation holds for multivariate time series data.

\section{Methods}

We examined how missing data mechanisms impact the recovery of parameter estimates, by performing a Monte-Carlo simulation. In this simulation, data was generated and a portion of the data was deleted to create MCAR, MAR, TMAR, ATMAR, and MNAR samples with $\beta$ coefficients from the corresponding logistic regressions found via grid search. The resulting data sets were analyzed in an idiographic discrete time continuous measure state-space model with Kalman filtering as a baseline method and compared with multivariate time series data imputation methods

\subsection{Data Generating Model}

The data generating model for this simulation is the idiographic discrete time state-space model with linearly related normally distributed states and normally distributed observations. This model can be viewed as a dynamic factor model. To keep the simulation simple, we simulated a 2 state, 3 indicators per state model of the form of equations 1-4, where $\mathbf{x}_{t}$ is the $2 \times T$ matrix of states $\mathbf{A}$ is the $2 \times 2$ matrix of transition coefficients, $\mathbf{0}_{p}$ is the $2 \times 1$ mean zero vector, $\mathbf{Q}$ is the $p \times p$ covariance matrix for the state error, $\mathbf{z}_t$ is the $6 \times 1$ vector of measurements at time t, $\mathbf{H}$ is the $6 \times 2$ matrix that maps states to measurements, and $\mathbf{R}_{t}$ is the $6 \times 6$ covariance matrix for measurement error. This produces a $T \times 8$ dataframe with 2 columns of T-many state values and 6 columns of T-many measured values.

\subsection{Conditions}

 Parameter values were chosen to mimic real conditions of small, medium, and large effect sizes and for the proportion of variance explained by the dynamic parameters. There were be 2 states with 3 items per state. The state error covariance $\mathbf{Q}$ is set to identity for all conditions. Measurement error is operationalized by choosing values in the error covariance matrix $\mathbf{R} =\begin{pmatrix}
\sigma^{2} & 0\\
0 & \sigma^{2}
\end{pmatrix}$ and the loading matrix $\mathbf{H}$ matrix is of the form $\begin{pmatrix} \lambda & \lambda & \lambda & 0 & 0 & 0 \\
0 & 0 & 0 & \lambda & \lambda  & \lambda
\end{pmatrix}$ such that the following equality holds: $\sigma^2 + \lambda^2 = 1$.  We varied the measurement error in 2 conditions: low measure error/high loadings ($\sigma^2 = .25, \lambda^2 = .75$) and high measurement error/low loadings ($\sigma^2 = .75, \lambda^2 = .25$) 

The $\mathbf{A}$ matrix (containing autoregressive and crosslagged effects) takes the form $\begin{pmatrix}
\alpha &  \gamma \\
0  &  \alpha
\end{pmatrix}$, where $\alpha$ $\in$ $\{$.2,.7$\}$ to examine differing strengths, small and large, of autoregressive relations and small, medium, and large $\gamma$ $\in$ $\{$0, .15, .3$\}$ (fully crossed) to examine how missingness mechanisms interact with cross-lagged relations (no relation between states, moderate relation between states, and strong relationship between states). The autoregressive effects explain about 4\% of the variance when equal to .2, and about 32\% of the variance when equal to .7. On the other hand, the cross-lagged effects explain about 2.5\% of the variance when equal to .15 and 8\% of the variance when equal to .3.

\subsubsection{MCAR}

For the MCAR mechanism, we simulated 2 levels of missingness: $15\%$ and $30\%$. All indicators for the state $\mathbf{x}_{1}$ with the targeted indices were set to NA, corresponding with all direct information about state $x_1$ being missing.

\subsubsection{MAR, TMAR, and ATMAR}

For the MAR data, a logistic regression was run with the two states as predictors with positive coefficients, then the probability of missingness was set to $$p(z_{1t},z_{2t},z_{3t} = NA) = \frac{1}{1 + \exp(\beta_{0}+\beta_{MAR}x_{2t})}$$ Indices of missingness were chosen based on the probabilities of missingness, $\beta_{0} \in \{4, 1.5\}$ and $\beta_{MAR} \in \{-3.5, -3\}$ for 15\% 30\% missingness at +1$\sigma^{2}$ above the mean of $x_2$, respectively. 

For TMAR data, we assumed there are 50 days with 10 time points per day, and an auxiliary variable $D \in \{1:10\}$ was calculated. A logistic regression for the probability of state $x_1$ indicators ($z_{1t},z_{2t},z_{3t}$) being missing as a function of $D$ is $$p(z_{1t},z_{2t},z_{3t} = NA) = \frac{1}{1 + \exp({\beta_{0} + \beta_{TMAR}D_t})}$$ with $\beta_{0} \in \{3, 2\}$ and $\beta_{TMAR} \in \{-.2, -.2\}$ for 15\% 30\% missingness for $D = 1$, respectively. $\beta$ values would need to be altered dependent on values of $D$.

For an ATMAR condition, we used a logistic function with the the probability of missingness set to $$p(z_{1t},z_{2t},z_{3t} = NA) = \frac{1}{1+\exp({\beta_{0} + \beta_{ATMAR}x_{1(t-1)}})}$$ with $\beta_{0} \in \{4, 1,5\}$ and $\beta_{ATMAR} \in \{-3.5, -3\}$ for 15\% 30\% missingness at $x_1 = 1$, respectively.  Note, this condition, while labeled as MAR, mixes MAR and MNAR mechanisms, as the values for $z_{1[t-1]},z_{2[t-1]},z_{3[t-1]}$ can be observed or missing.

\subsubsection{MNAR}

Recall that the difference between MAR and MNAR is that in MAR the missingness is dependent on other variables and, in MNAR, the missingness is dependent on the variable from which the data is missing. Thus, we used a similar model of generating missing data for MAR and MNAR. We set up an logistic regression with our variable of interest as our dependent variable and our variable of interest as our predictor, then, the probability of missingness was set to $$p(z_{1t},z_{2t},z_{3t} = NA) = \frac{1}{1+\exp({\beta_{0}+\beta_{MNAR}X_1t})}$$ with $\beta_{0} \in \{4, 1.5\}$ and $\beta_{MNAR} \in \{-3.5, -3\}$ 15\% and 30\% missingness at $X_1 = 1$, respectively.

\subsection{Missing data imputation}

The missing data imputation methods that were used were the Kalman filter \citep[\{dlm\} R package:][]{petris_r_2010}, various MICE conditons (see above MICE section) with the \citep[\{mice\} R package:][]{van_Buuren_2011}, and the EM algorithm with initializations of ARIMA, regression, and natural cubic spline \citep[\{mtsdi\} R package:][]{junger_imputation_2015}). Default settings were used in all cases except MICE-t, which uses a lagged matrix to predict the values of $\mathbf{z}$ at time $t$ from the values of $\mathbf{z}$ at time $t-1$ (as opposed to the Kalman filter which predicts time $t$ states from time $t-1$ states). For MICE-def and MICE-norm, the imputation models for $\mathbf{z}_1, \mathbf{z}_2$ and $\mathbf{z}_3$ were calculated using only $\mathbf{z}_4, \mathbf{z}_5$ and $\mathbf{z}_6$ (as each of $\mathbf{z}_1, \mathbf{z}_2, \mathbf{z}_3$ are set as missing simultaneously) at the same time point, and for MICE-t and MICE-tn, this pattern of prediction is used from the previous time point (i.e $\mathbf{z}_{[1-3]t}$ as a function of $\mathbf{z}_{[1-6]t-1}$).  With MICE-def, in order for $\mathbf{z}_4$-$\mathbf{z}_6$ to have use in imputation, they need to have to have cross-sectional predictive ability on $\mathbf{z}_{1}$-$\mathbf{z}_{3}$. This only occurs when $\gamma$ is non-zero, and, even then, the dependence would be weak. 

\subsection{Simulation overview}
A $2 (\text{measurement error conditions}) \times 2 (\alpha \text { conditions}) \times 3(\gamma \text { conditions})$  cell design was run for 100 replications for each cell to produce the raw timeseries (before missingness), with 500 timepoints per replication. The $10$ conditions of missingness mechanisms were then applied to the previously simulated raw data. For each of the 10 datasets with missing data per replication, the 8 missing data imputation methods were applied. For missing data imputation methods that generated multiple imputed datasets (i.e., MICE), 10 imputed datasets were generated, and parameter estimates/standard errors will be combined according to the standard practice of \citet{rubin_multiple_1996}. 

\subsection{Simulation Models and Outcomes}

Using the \{dlm\} R package \citep{petris_r_2010}, idiographic discrete time normally distributed state/measurement state-space models described in Equations 12-15 were fit, with the following free and fixed parameters: $\hat{\mathbf{A}} = \begin{pmatrix}
\hat{\alpha} &  \hat{\gamma} \\
0  &  \hat{\alpha}
\end{pmatrix}$, $\hat{\mathbf{H}}= \begin{pmatrix} \hat{\lambda}_1 & \hat{\lambda}_2 & \hat{\lambda}_3 & 0 & 0 & 0 \\
0 & 0 & 0 & \hat{\lambda}_4 & \hat{\lambda}_5  & \hat{\lambda}_6
\end{pmatrix}$, $\hat{\mathbf{R}} =\begin{pmatrix}
\hat{\sigma^{2}} & 0\\
0 & \hat{\sigma^{2}}
\end{pmatrix}$  and for identification purposes, the state error covariance matrix is fixed at $\mathbf{I}_2$. Parameter values were chosen based on the principle that they are not extreme values (i.e., they are neither too weak or too strong).

The following were computed for each cell for $\alpha$, $\gamma$, $\sigma^{2}$ and $\lambda^{2}$.

\subsubsection{Median bias}

$$\Delta \theta = \text{Median}(\theta - \hat{\theta})$$

where the $\theta$ are the true values of the parameter that are the assumed matrices and $\hat{\theta}$ are the estimated parameters. Bias is found, then the median of each condition is found.

\subsubsection{Median absolute relative bias}

$$\Delta\theta_{Rel} = \frac{|\theta - \hat{\theta}|}{\theta}$$

where the $\theta$ and $\hat{\theta}$ values are as above. Relative bias is found, then the median of each condition is found.

\subsubsection{Standard error and coverage}

The bias in standard error, $SE$, of estimates is contained in the Supplementary Materials.

Confidence intervals will be calculated as follows:

$$[\hat{\theta} - 1.96SE, \hat{\theta} + 1.96SE]$$

where $\overline{x}$ is the mean of a condition and $SE$ is defined as above. The coverage for a given parameter $\theta$ is the number of replications where the above confidence interval contains the true parameter $\theta$.

\section{Results}

We found the following general trends in the recovery of the autoregressive, cross-lagged, loadings, and measurement error parameters. First, as one might expect, there is greater bias with great percentage of missing. Second, the parameters associated with missingness (i.e. $\alpha_{11}$, $\gamma_{12}$, $\lambda_{11}-\lambda_{13}$, and $\sigma_{1}-\sigma_{3}$) show more bias than the complete parameters: if a parameter depends on missing data for its estimation, it shows more bias. Third, increasing the strength of the true parameters results in more bias. Fourth, the Kalman filter performed well for missing data for idiographic discrete time continuous measure state-space models, while the other methods struggled. Finally, the missingness mechanisms can be divided into two groups based on the amount of bias and variability they show with less bias and variability being associated with MCAR, MAR, and TMAR and more bias and variability associated with ATMAR, and MNAR. These trends are seen in each of the parameters' recovery.

Tables 1 and 2 in the Supplemental Materials are tables of the median bias for $\alpha_{11}$, $\sigma^{2}_{1}$, $\sigma^{2}_{2}$, $\sigma^{2}_{3}$, $\lambda_{11}^{2}$, $\lambda_{12}^{2}$, and $\lambda_{13}^{2}$ at 30\% missingness for the TMAR conditions $\gamma$ = 0 and $\gamma$ = .3, respectively, while Tables 3 and 4 are similarly for the ATMAR conditions. As we saw similar trends for MCAR, MAR, and TMAR, and ATMAR and MNAR, only TMAR and MNAR are shown. Notice that bias in $\alpha$ increases with the strength of the autoregressive effect. The $\sigma^{2}$s generally show low bias, while the bias in $\lambda^{2}$ increases with the strength of the loadings. The Kalman filter generally showed then least amount of bias (with the exception being when there is a low autoregressive effect for ATMAR and MNAR), while many of the MICE conditions showed larger amounts of bias. We see some increase in bias while increasing $\gamma$ and switching from TMAR missingness to ATMAR missingness. Note, the zeroes seen on the table were not true zeroes, but arose as a result of rounding to three decimal places.

Tables 5 and 6 in the Supplemental Materials are tables of the coverage for $\alpha$, $\gamma$ and $\lambda^{2}$ for TMAR missingness for $\gamma = 0$ and $\gamma = .3$, respectively, with the same arrangement for Tables 7 and 8 in the Supplemental Materials for ATMAR missingness. Notice that $\gamma$ falls into the confidence interval typical for low values of $\gamma$, though the success rate decreases with a higher value of $\gamma$. The success of recovering $\alpha$ and $\lambda^{2}$ is dependent upon imputation method and missingness mechanism. For both parameters, the EM-Regression imputation method performed the worst. For $\alpha$ with no cross-lagged effects, the Kalman filter with TMAR missingness performed the best. For $\lambda^{2}$ with no cross-lagged effects, the EM-ARIMA imputation method with TMAR missingness performed the best. Similar trends held for $\alpha$ and $\lambda^{2}$ with strong cross-lagged effects.

\subsection{State Autoregression Parameters ($\alpha$)}

\begin{figure}[H]
\centering
\includegraphics[width=\textwidth]{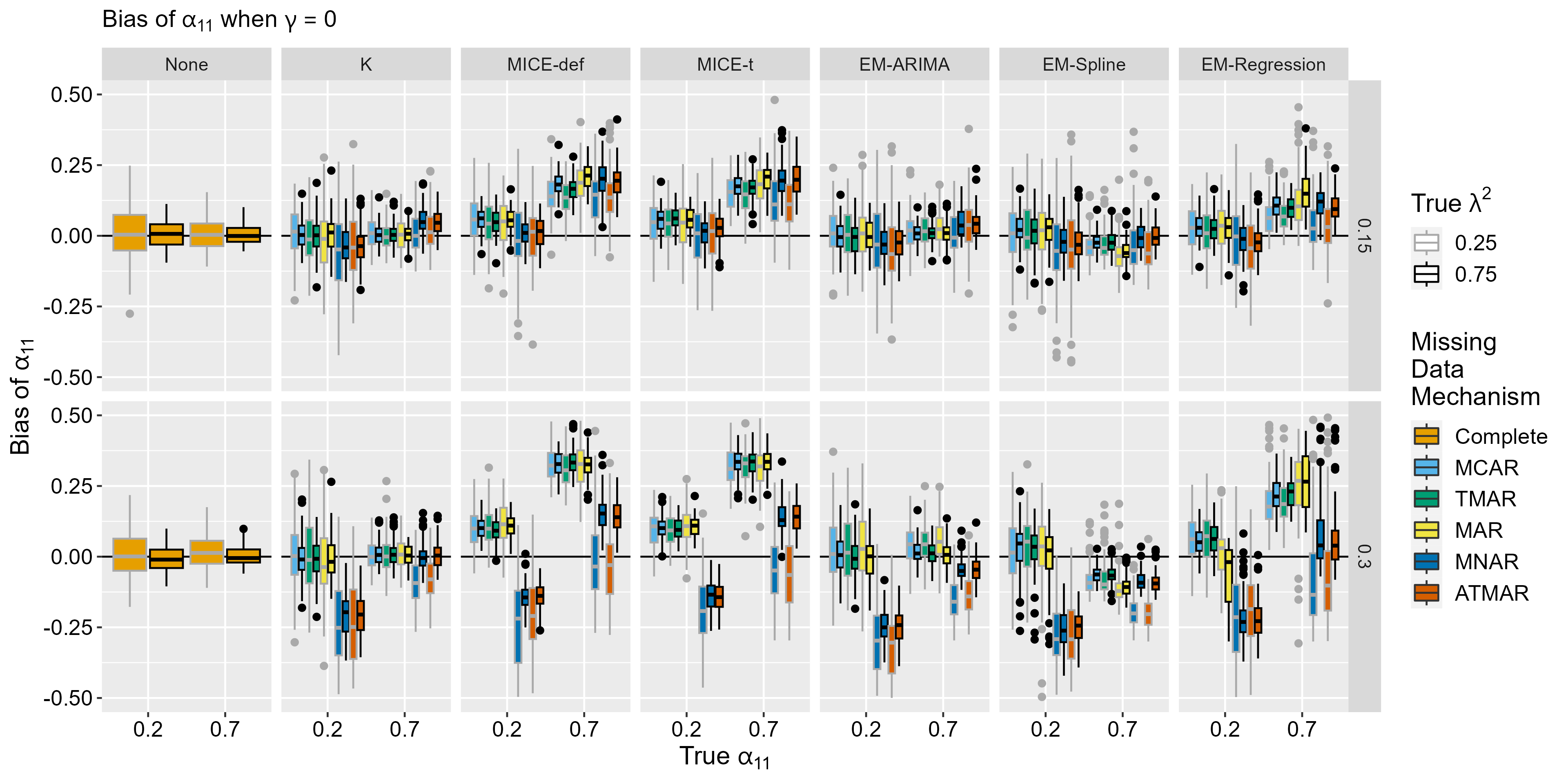}
\caption{The above graphs shows the bias of the estimated $\alpha$ (on the y-axis) by the true $\alpha$ (on the x-axis for $\gamma$ = 0. The respective missingness mechanisms are shown: complete data (orange), MCAR (sky blue), TMAR (green), MAR (yellow), MNAR (dark blue), and ATMAR (dark orange). The outlines on the box plots show the differing loading conditions: light grey for low loadings/high measurement error and black for high loadings/low measuremenent error. As the y-axis was restricted to range from -.5 to .5, 926 outliers were removed, primarily from the EM-Regression condition. For the box plots, the bottom of the box is the first quartile, the central line is the median, the top of the box is the third quartile, the whiskers extend 1.5(Interquartile Range), and the dots beyond the whiskers are outliers.}
\end{figure}

\begin{figure}[H]
\centering
\includegraphics[width=\textwidth]{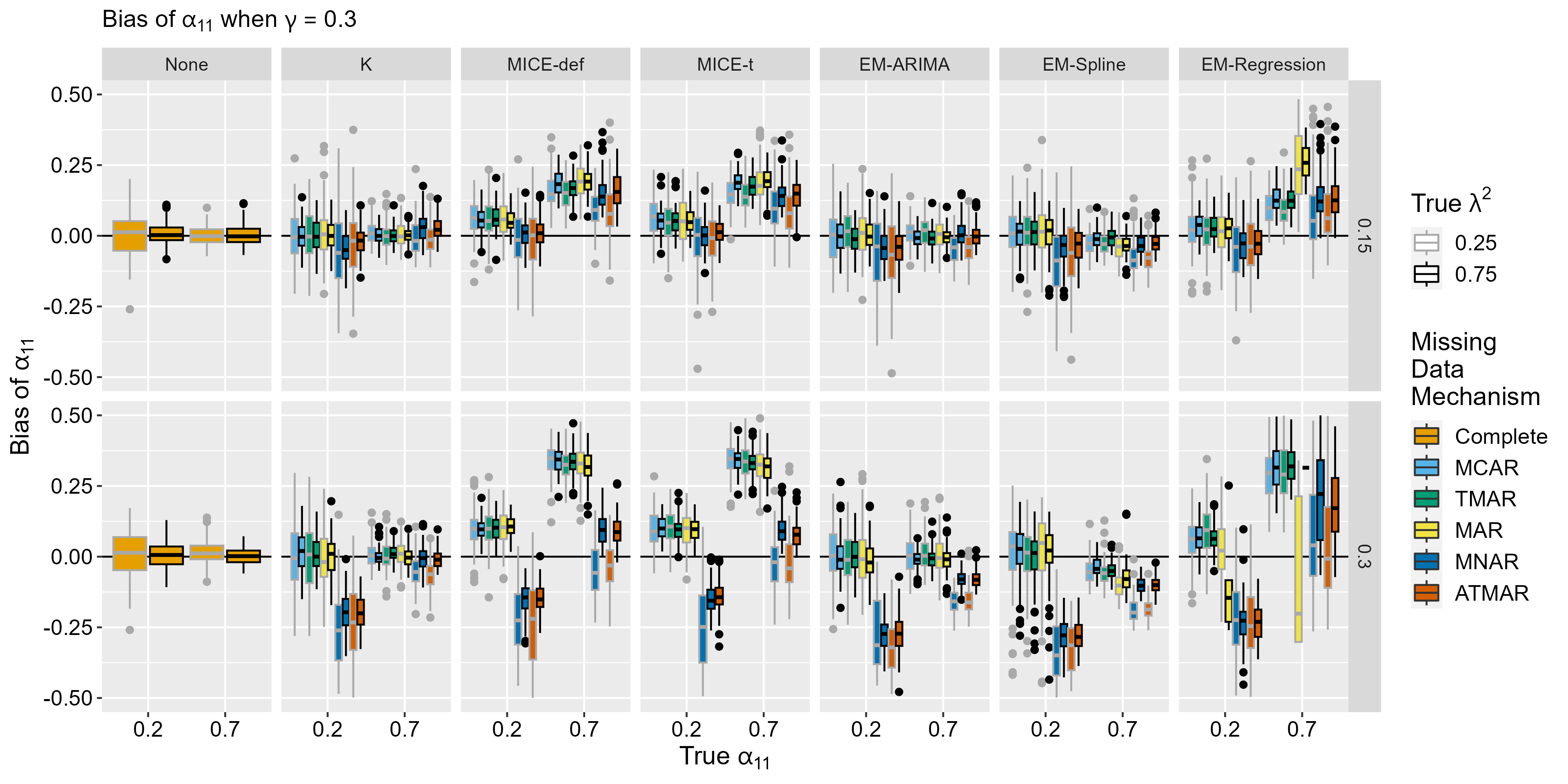}
\caption{The above graphs shows the bias of the estimated $\alpha$ (on the y-axis) by the true $\alpha$ (on the x-axis for $\gamma$ = .3. The respective missingness mechanisms are shown: complete data (orange), MCAR (sky blue), TMAR (green), MAR (yellow), MNAR (dark blue), and ATMAR (dark orange). The outlines on the box plots show the differing loading conditions: light grey for low loadings/high measurement error and black for high loadings/low measuremenent error. As the y-axis was restricted to range from -.5 to .5, 1171 outliers were removed, primarily from the EM-Regression condition. For the box plots, the bottom of the box is the first quartile, the central line is the median, the top of the box is the third quartile, the whiskers extend 1.5(Interquartile Range), and the dots beyond the whiskers are outliers.}
\end{figure}

We found that $\alpha_{11}$, the autoregressive parameter associated with missing data, had greater bias than $\alpha_{22}$, the autoregressive parameter associated with complete data, though in Figures 4 and 5 we only include $\alpha_{11}$. These figures show the bias of the $\alpha_{11}$ parameter estimates with respect to $\lambda^2$ and missingness mechanisms, for $\gamma$ = 0 and $\gamma$ = .3 respectively. For all graphs, the trend in $\gamma$ was linear (i.e., the relationship between $\gamma$ = 0 and $\gamma$ = .3 is merely a more extreme version of the relationships between $\gamma$ = 0 and $\gamma$ = .15, and $\gamma$ = .15 and $\gamma$ = .3), so, for all of the graphs, we only consider the $\gamma$ = 0 and $\gamma$ = .3 conditions. In Figure 4, we see that there is no effect of $\lambda^{2}$, but bias increases as the true value of the autoregressive effect increases. There is greater bias when increasing the amount of missingness. The Kalman filter is the most successful method for recovering the autoregressive effects for all conditions except when there is MNAR or ATMAR missingness with a weak autoregressive effect, while the multiple imputation methods struggle. Notice that the MCAR graph is similar to the TMAR graph and the MAR graph in that they have less bias and variability than the ATMAR graph which is similar to the the MNAR graph in that they have greater bias and variability. In Figure 5, for the conditions that struggle to recover $\alpha$ (i.e., EM-regression with high autoregressive effects), there is an effect of $\gamma$, otherwise there is not; however, as the strength of the true autoregressive effect increases, the bias increases. With an increase in percent of missingness comes an increase in bias. Again, the Kalman filter succeeds (except in the same condition as above), where the multiple imputation methods fail, and the ATMAR/MNAR data shows greater bias and variability. We do not see an effect of $\gamma$.

\subsection{State Cross-Lag Parameter ($\gamma$)}

\begin{figure}[H]
\centering
\includegraphics[width=\textwidth]{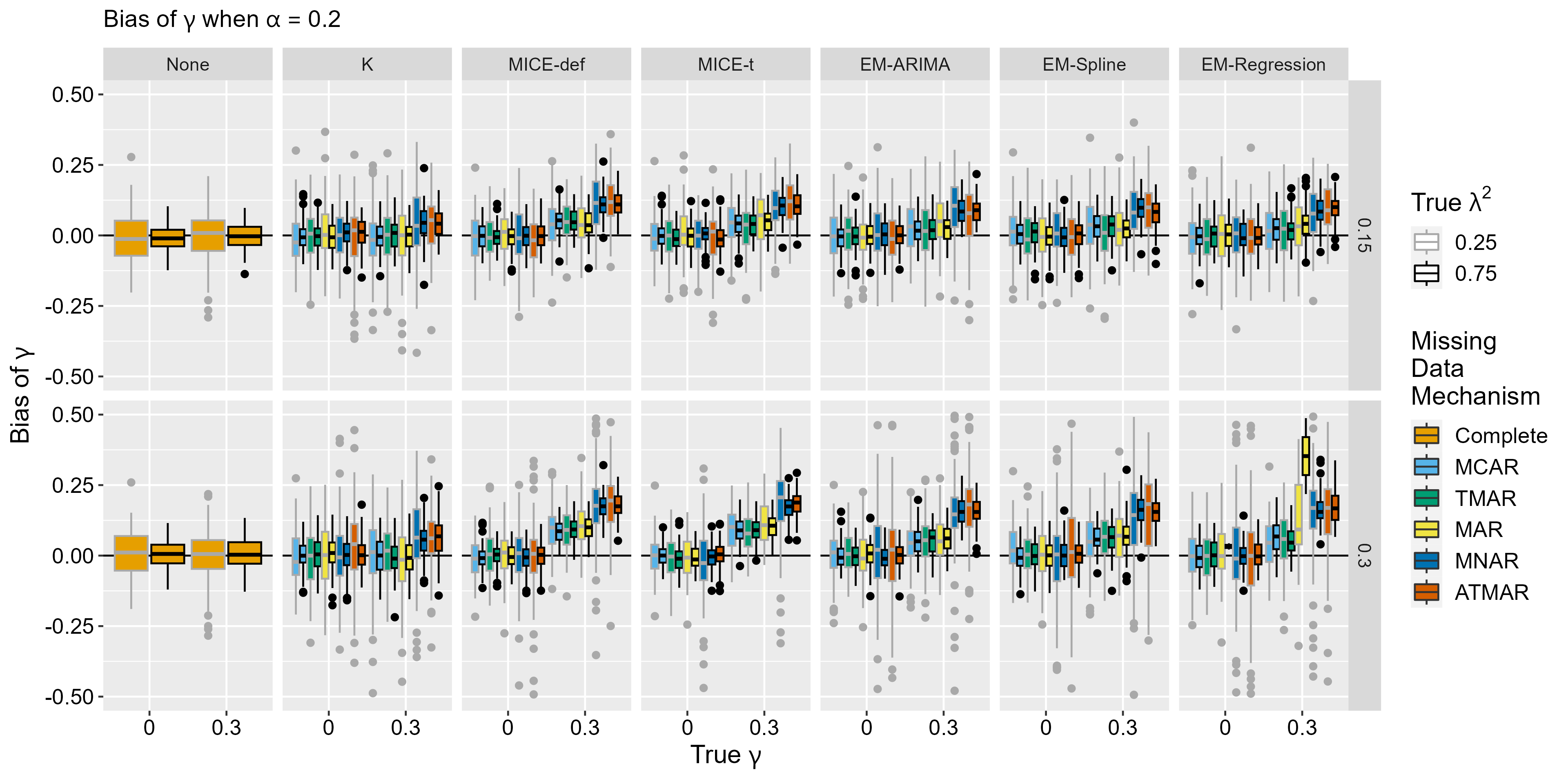}
\caption{The above graphs shows the bias of the estimated $\gamma$ (on the y-axis) by the true $\gamma$ (on the x-axis for $\alpha$ = .2. The respective missingness mechanisms are shown: complete data (orange), MCAR (sky blue), TMAR (green), MAR (yellow), MNAR (dark blue), and ATMAR (dark orange). The outlines on the box plots show the differing loading conditions: light grey for low loadings/high measurement error and black for high loadings/low measuremenent error. As the y-axis was restricted to range from -.5 to .5, 440 outliers were removed, primarily from the EM-Regression condition. For the box plots, the bottom of the box is the first quartile, the central line is the median, the top of the box is the third quartile, the whiskers extend 1.5(Interquartile Range), and the dots beyond the whiskers are outliers.}
\end{figure}

\begin{figure}[H]
\centering
\includegraphics[width=\textwidth]{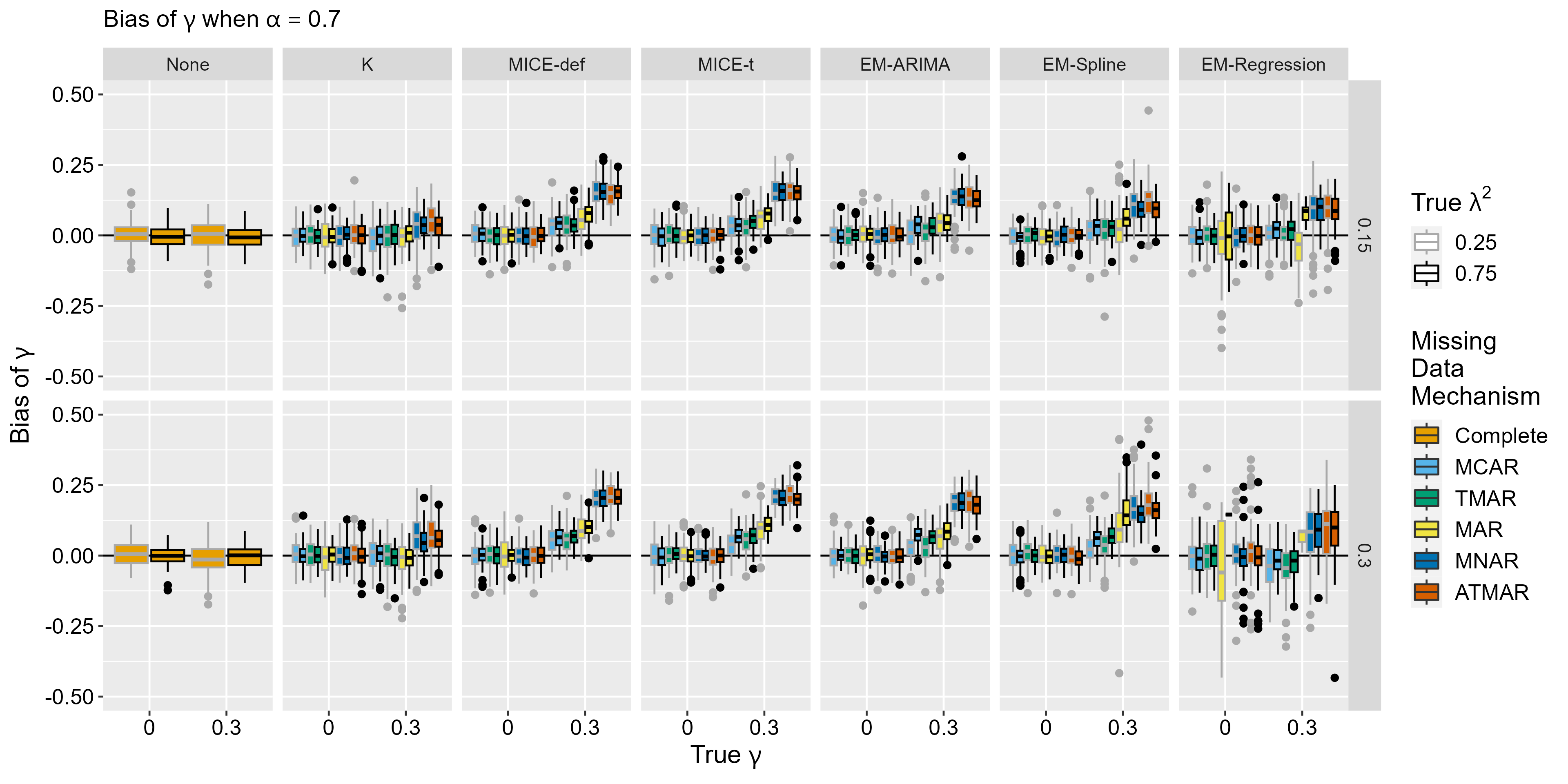}
\caption{The above graphs shows the bias of the estimated $\gamma$ (on the y-axis) by the true $\gamma$ (on the x-axis for $\alpha$ = .7. The respective missingness mechanisms are shown: complete data (orange), MCAR (sky blue), TMAR (green), MAR (yellow), MNAR (dark blue), and ATMAR (dark orange). The outlines on the box plots show the differing loading conditions: light grey for low loadings/high measurement error and black for high loadings/low measuremenent error. As the y-axis was restricted to range from -.5 to .5, 748 outliers were removed, primarily from the EM-Regression condition. For the box plots, the bottom of the box is the first quartile, the central line is the median, the top of the box is the third quartile, the whiskers extend 1.5(Interquartile Range), and the dots beyond the whiskers are outliers.}
\end{figure}

We found that $\gamma_{12}$, the cross-lag relation from $\mathbf{x}_1$ (the variable with missing indicators) to $\mathbf{x}_2$. had more bias that $\gamma_{21}$, the $\gamma$ associated with complete data (and set to null), though Figures 6 and 7 only depict $\gamma_{12}$. These figures show the bias of the $\gamma_{12}$ parameter with respect to $\lambda^{2}$ and missingness mechanisms for $\alpha$ = .2 and $\alpha$ = .7 respectively. In Figure 6, we see that the bias increases with an increase in the strength of the true $\gamma$ parameter. Increasing the percent missingness increases bias. $\gamma$ is recovered well for MCAR, MAR, and TMAR, but there is more variability and bias in ATMAR and MNAR. Again, we see that the Kalman filter excels at recovering the $\gamma$ parameter, while the MICE conditions struggle. In Figure 7, we see increasing the strength of the true $\gamma$ parameter increases the bias. With an increase in percent missingness, comes an increase in bias. MCAR, MAR, and TMAR are successful at recovering the $\gamma$ parameter, while ATMAR and MNAR show greater bias and variability. Finally, the Kalman filter is again the best method for handling missing data, while MICE-def  shows more bias. We do not see an effect of $\alpha$.

\subsection{Measurement Loadings and Error ($\lambda^{2}$ and $\sigma^{2}$)}

For $\sigma^2$, we saw similar bias between $\sigma^{2}_{1}$, $\sigma^{2}_{2}$, and $\sigma^{2}_{3}$, the $\sigma^{2}$s associated with missingness, which was higher than the similar biases $\sigma_{4}$, $\sigma_{5}$, and $\sigma_{6}$, the $\sigma^{2}$s associated with complete data. Due to this similarity, $\sigma^{2}_{1}$, $\sigma^{2}_{2}$, and $\sigma^{2}_{3}$ are treated as one variable in Figures 1 and 2 in the supplementary materials. These graphs show the bias of $\sigma^{2}$ with respect to $\lambda^2$ and missingness mechanisms for $\alpha$ = .2 and $\alpha$ = .7 respectively. In Figure 1 in the supplementary materials, there is greater variability with greater measurement error. Increasing percent missingness increases bias. We see lesser bias and variability in MCAR, MAR, and TMAR and greater bias and variability in ATMAR and MNAR. Unlike in the previous cases, the Kalman filter and the MICE conditions can all recover the $\sigma^{2}$ parameters. In Figure 2 in the supplementary materials,  again we see that increasing the strength of the measurement error and the percentage of missingness results in greater bias. MCAR, MAR, and TMAR show lesser bias and variability, while ATMAR and MNAR show greater bias and variability. Again, we see the contrary result that, in addition to the Kalman filter and both MICE conditions  can recover the $\sigma^{2}$ parameter. For Figures 1 and 2 in the supplementary materials, we do not see an effect of $\alpha$.

For $\lambda^2$, we saw similar bias between $\lambda_{11}^{2}$, $\lambda_{12}^{2}$, and $\lambda_{13}^{2}$, the $\lambda^{2}$s associated with missingness, which was higher than the similar biases $\lambda_{24}$, $\lambda_{25}$, and $\sigma_{26}$, the $\sigma^{2}$s associated with complete data. Due to this similarity, $\lambda_{11}^{2}$, $\lambda_{12}^{2}$, and $\lambda_{13}^{2}$ are treated as one variable in Figures 3 and 4 in the supplementary materials . These graphs show the bias of $\lambda^{2}$ with respect to $\gamma$ and missingness mechanisms for $\alpha$ = .2 and $\alpha$ = .7 respectively. In figure 3 in the supplementary materials, we see no effect of $\gamma$, though increasing the strength of the loadings and the percent of missingness results in greater bias. ATMAR and MNAR have greater variability and bias than MCAR, MAR, and TMAR. For the lower bias conditions (i.e., MCAR, MAR and TMAR), the Kalman filter and the MICE conditions recover the loadings successfully, but struggle with MNAR and ATMAR conditons. In Figure 4 in the supplementary materials, we also see increased bias for the $\lambda^{2}$s associated with missingness as compared to the $\lambda^{2}$s associated with complete data. We see no effect of $\gamma$, though increasing the strength of the loadings and the percentage of missingness increases bias. We see the same missingness groupings as MCAR, MAR, and TMAR are associated with lesser bias and variability and ATMAR and MNAR are associated with greater bias and variability. Finally, Kalman filter again recovers the $\lambda^{2}$ parameter, while both MICE conditions fail to successfully recover the $\lambda^{2}$ parameter. Comparing the two graphs, we see an effect of $\alpha$ in that there is greater bias and variability with stronger $\alpha$.

\section{Discussion}

Overall, we offer the following recommendations for handling missing EMA data in idiographic discrete time continuous measure state-space models. The Kalman filter is a good choice for missing data imputation for MCAR, MAR, or TMAR: the approach resulted in the smallest parameter bias in these conditions. Furthermore, with a large autoregressive effect, the Kalman filter is also recommended for MNAR and ATMAR missingness. With the default settings, multiple imputation methods (i.e., MICE-def and MICE-t) struggle to recover the autoregressive and cross-lagged effects, with bias in parameter recovery particularly high in MNAR and ATMAR settings. Finally, If the missingness is MCAR, MAR, or TMAR as opposed to ATMAR or MNAR, there will be less issues with parameter estimation.

First, it was expected that the bias would be greater for the parameters associated with missingness than the parameters associated with complete data and for stronger true parameters. These former parameters are more impacted by missing observations whereas the latter parameters rely on complete, unaltered information. Thus, we expect to see more bias in these conditions. Second, it also makes sense that as the strength of the parameters increases, the bias increases. However, particularly for the estimates of the autoregressive and cross-lagged parameters, larger magnitude effects led to improved performance of the Kalman filtering approach. Third, it is expected that increased missingness results in increased bias. In these conditions, there are simply fewer of the true values, allowing for more opportunities for biased estimates.

Fourth, the Kalman filtering approach performed best out of all the methods except for MNAR and ATMAR missingness with a low autoregressive effect. Recall that the Kalman filter estimates the expected values of states using both previous values of states and observed measurements. As such, the Kalman filter (and, also, other filtering methods) explicitly works on the level of unobserved states, while the other imputation methods are all attempting to impute the missing observed variables. This explains the improved performance of the Kalman filter for strong autoregressive and cross-lagged relations. When the states are strongly linked in time, that means your prediction of the states at a timepoint with missing data will be better (because you know that the states are very similar to previous timepoints). However, generally MNAR and ATMAR missingness are more challenging conditions for parameter recovery. This paired with a low autoregressive effect results in more bias than other conditions due to the Kalman filter relying on having a reasonably strong autoregressive effect, as that means it can carry forward information to fill in the missingness. Thus, in this condition, not only is there have a weak temporal dependency, but there is also high missingness which makes it even more difficult to obtain good estimates. Additionally, an important limitation here is that we knew the true data generating process was a state-space model, which has implications for how the observed data is related to the unobserved states. Given that, it makes sense that the Kalman filter is performing well in this setting, but, for other time series models and in empirical settings, the performance of the Kalman filter will likely be worse than what is observed here.

Fifth, it was a surprise that MICE-def could not recover the autoregressive and cross-lagged parameters, but performed well with the measurement error parameter. Multiple imputation is generally the recommended method for handling missing data \citep{rubin_multiple_1996}, so it is unexpected that it performed poorly. Furthermore, we used the default settings in the functions as we expect most users to do the same. Because of the general performance of multiple imputation with the default settings, we do not recommend using these packages for imputation with the default settings. Even so, recall what MICE-def does: it is a cross-sectional method which builds the imputation model without considering temporal relations. Hence, it makes sense that it can recover the measurement error (as this is purely cross-sectional), but struggles with recovering the dynamics. In an attempt to resolve this issue, the MICE-t condition did take into account the temporal relations in the data with a lag-1 matrix serving as the predictor in the imputation model. Even with this alteration, MICE-t struggled in ways similarly to  MICE-def. We hypothesize that, though MICE-t does take into account the time dependence, it still performs less well than the Kalman filter because it cannot take into account information from the latent states. 

Sixth, the EM algorithms, particularly EM-ARIMA and EM-Spline performed well with lower levels of missing data (i.e., 15\% missingness), though struggled with higher levels of missingness (i.e., 30\% missingness). EM-Regression generally performed poorly, and, from many of the analyses for this method, we had to remove outliers. Because the success of the EM-Regression method depends on the predictive relationship between $\mathbf{z}_{4}$-$\mathbf{z}_{6}$ and $\mathbf{z}_{1}$-$\mathbf{z}_{3}$, and given that this varies with $\gamma$ (in that low values of $\gamma$ would correspond to less predictive power), it makes sense that this method struggled. In addition, the EM algorithm is an algorithm for obtaining maximum likelihood estimates, and it is well-established that using these estimates to generate imputations has numerous shortcomings \citep{gomez-carracedo_practical_2014}. Such a method amounts to regression imputation (in that the expectation step is simply the results of predictions from a series of regression equations), resulting in restricted variation, biased associations, among other problems \citep{gomez-carracedo_practical_2014}. Despite EM-ARIMA and EM-Spline performing well at lower levels of missingness, using maximum likelihood estimation followed by single imputation is not recommended as a missing data solution.

Finally, we see a division between MCAR/MAR/TMAR and ATMAR/MNAR. Recall that MCAR and TMAR are both missing data mechanisms that do not rely on other variables in the model for their missingness. Hence, it is expected that they would perform similarly. However, we reiterate the warning that most if not all missing data will not be MCAR (and will rarely be purely TMAR). Additionally, MNAR and ATMAR performed approximately the same (notably with Kalman filtering resulting in the lowest bias for high autoregressive effects). This is likely because MNAR and ATMAR are two aspects of the same type of missingness mechanisms. Where MNAR is missingness based on the missing variable's value, ATMAR is missingness based on a previous timepoint's value (which may or may not be missing). This results in a danger, as not taking into account temporal dependency might result in reduced model performance, and an opportunity, as temporal dependency offers more information to impute missing values from. Further research on missingness mechanisms and how they unfold across time should be pursued.

The reader may having the following concerns about this project: the use of default methods for MICE-def, no varying of states and indicators or number of time points, and what if the phenomenon is continuous.  First, we used the default settings for MICE-def as this will be the most common choice by users of the software, and that the default settings (with use predictive mean matching) have been shown to be effective across a number of settings. Due to the lack of success with MICE-def, we followed up with MICE-t, the time-dependent imputation mode, M-norm, the Bayesian linear regression method, and M-tn, the combination of the two. We found that, even with improvements to MICE, the method was unable to recover the time-series parameters that were dependent on states. Second, we did not increase the number of states or indicators, as we expect the bias and issues to increase with an increasing number of states, and we expect the measurement to improve with an increasing number of indicators (which will likely result in lower bias, but not necessarily for the multiple imputation methods). Third, we evaluated only one sample size (500 time points) as we assume that any bias due to missingness will be exacerbated with smaller numbers of time points. Fourth, there are limits to imputation. If you are modeling a continuous process as discrete, you are missing all of the infinitesimal points between the discrete time points. There are limits to how much missing data can be handled by an imputation method (see below), and this would be a case where it would be best to just use a continuous time model. 

Taken together, these results suggest several critical considerations when accounting for missingness in EMA settings. First, the relatively good performance of Kalman filtering relative to other methods suggest that making use of the temporal information in intensive longitudinal data is critical. If there is a strong autoregressive effect, this implies that the imputation of missing values at time $t$ would benefit from including information from time $t-1$. Additionally, if there is a latent structure (as we simulated here), then taking advantage of that structure will improve the performance of any imputation method. This is due to the presence/absence of measurement error, and the Kalman filter, in this case of a correctly specified model, likely benefited from accounting for measurement error in its imputation of the missing data. However, as was noted previously, while the performance of the Kalman filter benefited from accounting for these features, when these features were less apparent in the data generating process (i.e. low autoregressive effects, high amounts of missingness), then the performance suffered relative to MICE based imputation methods. These results reify a fundamental property of missing data methods: the more relevant structural information integrated into the missing data model, the better the missing data model will perform. In the setting of EMA data, particularly EMA data of psychological constructs, this suggests that both temporal dependency and measurement error are key features to address in any missingness model.

For future directions, first, we would like to further evaluate the use of multiple imputation methods for time series and state-space data as this analysis relied on the default setting and, also, include Amelia in our analyses. These methods are flexible, which implies that, with different base modeling choices, we could theoretically improve performance. However, this would be specific to different datasets and models. Second, it is often recommended that the way to handle missing data in a state-space model is to fit a continuous time model. As a next step, we will compare the best discrete time missing data method, the Kalman filter, to a continuous time model. For a continuous time model, a Kalman-Bucy filter is used which discretizes the continuous time into small intervals \citep{Klmn1961NewRI}. If the underlying process is discrete, then this would simplify into the Kalman. However, if it is not, we expect to see differences in the discrete and continuous time models. Finally, we see that the Kalman filter is generally successful at 30\% missingness, but we are interested to see what the limit for amount of missingness is for the Kalman filter in order to provide better recommendations. Thus, we will be testing increasing amounts of missingness to see when the Kalman filter fails.

\bibliography{Bib1.bib}
\bibliographystyle{apacite}

\newpage
\setcounter{section}{0}

\end{document}


\title{Supplemental Materials for ``\textbf{Missing Data in Discrete Time State-Space Modeling of Ecological Momentary Assessment Data: A Monte-Carlo Study of Imputation Methods}''}
\maketitle

\section{Rubin's  equations for calculating the variance of the multiply imputed data}

Consider the complete data statistics, $Q$, and associated variance-covariance matrices, $\mathbf{U}$, with $Q_{*n}$ and $U_{*n}$ corresponding to particular imputed datasets of $m$ datasets. The repeated-imputation estimate is defined as $$\bar{Q}_{m} = \sum_{l = 1}^{m}\frac{Q_{*l}}{m}$$ with the corresponding variance-covariance matrix of $\bar{Q}_{m}$ defined as $$\mathbf{T}_{m} = \bar{U}_{m} + \frac{m+1}{m}B_{m}$$ where $\bar{U}_{m}$ is the within-imputation variability defined as $$\sum_{l=1}^{m}\frac{U_{*l}}{m}$$ and $B_{m}$ is the between-imputation variability defined as $$\sum_{l=1}^{m}\frac{(Q_{*l}-\bar{Q}_{m})(Q_{*l}-\bar{Q}_{m})'}{m-1}$$ 

\newpage

\section{Median Bias}

\begin{singlespace}
\begin{table}[H]
\centering
\caption{Median bias for $\alpha_{11}$, $\sigma^{2}_{1}$, $\sigma^{2}_{2}$, $\sigma^{2}_{3}$, $\lambda_{11}^{2}$, $\lambda_{12}^{2}$, and $\lambda_{13}^{2}$ when the missingness mechanism is TMAR, missing percentage is 30\%, and true $\gamma$ = 0. MCAR and MAR missingness behave similarly to TMAR missingness and were, thus, excluded.}
\begin{tabular}{cccccccccc}
\hline
Imputation    & True $\alpha$ & True $\lambda^{2}$ & $\alpha_{11}$ & \multicolumn{1}{c|}{$\sigma^{2}_{1}$} & $\sigma^{2}_{2}$ & $\sigma^{2}_{3}$ & $\lambda^{2}_{11}$ & $\lambda^{2}_{22}$ & $\lambda^{2}_{13}$ \\ \hline
MICE-tn       & 0.2           & 0.75               & 0.101         & 0.005                                 & -0.005           & 0                & -0.018             & -0.023             & -0.012             \\
MICE-tn       & 0.7           & 0.75               & 0.338         & -0.002                                & 0.006            & 0.001            & -0.468             & -0.477             & -0.475             \\
MICE-tn       & 0.2           & 0.25               & 0.111         & 0                                     & -0.001           & 0.001            & 0.001              & 0.006              & -0.015             \\
MICE-tn       & 0.7           & 0.25               & 0.325         & 0.018                                 & -0.002           & 0.015            & -0.19              & -0.142             & -0.174             \\
MICE-t        & 0.2           & 0.75               & 0.095         & -0.001                                & -0.001           & 0.002            & 0.016              & -0.002             & 0.002              \\
MICE-t        & 0.7           & 0.75               & 0.337         & 0.005                                 & 0.006            & -0.001           & -0.454             & -0.461             & -0.461             \\
MICE-t        & 0.2           & 0.25               & 0.093         & 0.024                                 & 0.009            & -0.002           & 0.006              & -0.009             & 0.005              \\
MICE-t        & 0.7           & 0.25               & 0.33          & -0.01                                 & 0.007            & 0.003            & -0.148             & -0.152             & -0.166             \\
MICE-norm     & 0.2           & 0.75               & 0.1           & 0.004                                 & 0.001            & -0.001           & -0.022             & -0.019             & -0.008             \\
MICE-norm     & 0.7           & 0.75               & 0.343         & 0.002                                 & -0.003           & -0.005           & -0.467             & -0.492             & -0.493             \\
MICE-norm     & 0.2           & 0.25               & 0.099         & -0.001                                & -0.005           & 0.005            & -0.009             & 0.003              & -0.004             \\
MICE-norm     & 0.7           & 0.25               & 0.319         & 0.01                                  & -0.002           & 0.008            & -0.159             & -0.153             & -0.158             \\
MICE-def      & 0.2           & 0.75               & 0.092         & 0.008                                 & 0.005            & -0.005           & 0.005              & 0.012              & 0.01               \\
MICE-def      & 0.7           & 0.75               & 0.333         & 0.004                                 & 0                & 0.009            & -0.433             & -0.452             & -0.454             \\
MICE-def      & 0.2           & 0.25               & 0.099         & 0.004                                 & 0.017            & 0.001            & -0.015             & -0.011             & -0.001             \\
MICE-def      & 0.7           & 0.25               & 0.306         & 0.017                                 & -0.004           & 0.015            & -0.159             & -0.147             & -0.159             \\
K             & 0.2           & 0.75               & -0.008        & 0.002                                 & 0.001            & -0.002           & 0.016              & 0.015              & 0.005              \\
K             & 0.7           & 0.75               & 0.007         & 0.004                                 & 0.007            & 0.005            & 0.001              & 0.002              & 0.013              \\
K             & 0.2           & 0.25               & -0.011        & 0.016                                 & 0.014            & 0.014            & -0.001             & 0.012              & 0.019              \\
K             & 0.7           & 0.25               & 0.001         & -0.002                                & 0.007            & 0.017            & 0.011              & 0.006              & 0.007              \\
EM-Spline     & 0.2           & 0.75               & 0.035         & 0.066                                 & 0.069            & 0.076            & 0.229              & 0.229              & 0.224              \\
EM-Spline     & 0.7           & 0.75               & -0.067        & 0.05                                  & 0.042            & 0.046            & 0.204              & 0.194              & 0.199              \\
EM-Spline     & 0.2           & 0.25               & 0.053         & 0.228                                 & 0.23             & 0.236            & 0.076              & 0.074              & 0.072              \\
EM-Spline     & 0.7           & 0.25               & -0.098        & 0.183                                 & 0.185            & 0.179            & 0.091              & 0.094              & 0.101              \\
EM-Regression & 0.2           & 0.75               & 0.063         & 0.067                                 & 0.074            & 0.069            & 0.194              & 0.185              & 0.197              \\
EM-Regression & 0.7           & 0.75               & 0.231         & 0.072                                 & 0.074            & 0.063            & -0.081             & -0.084             & -0.072             \\
EM-Regression & 0.2           & 0.25               & 0.071         & 0.206                                 & 0.196            & 0.199            & 0.067              & 0.069              & 0.075              \\
EM-Regression & 0.7           & 0.25               & 0.186         & 0.195                                 & 0.203            & 0.198            & -0.004             & -0.005             & -0.01              \\
EM-ARIMA      & 0.2           & 0.75               & -0.008        & 0.077                                 & 0.074            & 0.076            & 0.219              & 0.218              & 0.226              \\
EM-ARIMA      & 0.7           & 0.75               & 0.012         & 0.054                                 & 0.051            & 0.059            & 0.141              & 0.149              & 0.139              \\
EM-ARIMA      & 0.2           & 0.25               & 0.014         & 0.222                                 & 0.24             & 0.214            & 0.074              & 0.07               & 0.087              \\
EM-ARIMA      & 0.7           & 0.25               & 0.044         & 0.216                                 & 0.213            & 0.202            & 0.055              & 0.055              & 0.058             
\end{tabular}%

\label{tab:my-table}
\end{table}
\end{singlespace}

\newpage
\begin{singlespace}
\begin{table}[H]
\centering
\caption{Median bias for $\alpha_{11}$, $\sigma^{2}_{1}$, $\sigma^{2}_{2}$, $\sigma^{2}_{3}$, $\lambda_{11}^{2}$, $\lambda_{12}^{2}$, and $\lambda_{13}^{2}$ when the missingness mechanism is TMAR, missing percentage is 30\%, and true $\gamma$ = 0.3. MCAR and MAR missingness behave similarly to TMAR missingness and were, thus, excluded.}
\label{tab:my-table}
\begin{tabular}{cccccccccc}
Imputation    & True $\alpha$ & True $\lambda^{2}$ & $\alpha_{11}$ & $\sigma^{2}_{1}$ & $\sigma^{2}_{2}$ & $\sigma^{2}_{3}$ & $\lambda^{2}_{11}$ & $\lambda^{2}_{12}$ & $\lambda^{2}_{13}$ \\ \hline
MICE-tn       & 0.2           & 0.75               & 0.094         & 0.005            & 0                & -0.004           & -0.059             & -0.06              & -0.061             \\
MICE-tn       & 0.7           & 0.75               & 0.34          & 0.002            & 0.002            & 0.003            & -0.804             & -0.809             & -0.821             \\
MICE-tn       & 0.2           & 0.25               & 0.103         & 0.011            & 0.002            & 0.017            & -0.009             & -0.015             & -0.034             \\
MICE-tn       & 0.7           & 0.25               & 0.352         & -0.012           & 0.011            & -0.016           & -0.273             & -0.285             & -0.282             \\
MICE-t        & 0.2           & 0.75               & 0.096         & 0                & 0.003            & 0.004            & -0.031             & -0.049             & -0.03              \\
MICE-t        & 0.7           & 0.75               & 0.331         & -0.001           & -0.004           & 0.002            & -0.759             & -0.739             & -0.771             \\
MICE-t        & 0.2           & 0.25               & 0.107         & 0.021            & 0.022            & 0.02             & -0.006             & -0.003             & -0.024             \\
\end{tabular}
\end{table}
\end{singlespace}

\begin{singlespace}
\begin{table}[]
\centering
\caption{Median bias for $\alpha_{11}$, $\sigma^{2}_{1}$, $\sigma^{2}_{2}$, $\sigma^{2}_{3}$, $\lambda_{11}^{2}$, $\lambda_{12}^{2}$, and $\lambda_{13}^{2}$ when the missingness mechanism is ATMAR, missing percentage is 30\%, and true $\gamma$ = 0. MNAR missingness behaves similarly to ATMAR missingness and was, thus, excluded.}
\label{tab:my-table}
\begin{tabular}{cccccccccc}
Imputation    & True $\alpha$ & True $\lambda^{2}$ & $\alpha_{11}$ & $\sigma_{1}^{2}$ & $\sigma_{2}^{2}$ & $\sigma_{3}^{2}$ & $\lambda_{11}^{2}$ & $\lambda_{12}^{2}$ & $\lambda_{13}^{2}$ \\ \hline
MICE-tn       & 0.2           & 0.75               & -0.131        & -0.004           & -0.003           & 0                & 0.191              & 0.188              & 0.186              \\
MICE-tn       & 0.7           & 0.75               & 0.145         & 0.002            & -0.003           & 0.002            & -0.056             & -0.083             & -0.068             \\
MICE-tn       & 0.2           & 0.25               & -0.301        & -0.034           & -0.057           & -0.052           & 0.14               & 0.138              & 0.136              \\
MICE-tn       & 0.7           & 0.25               & -0.014        & -0.047           & -0.046           & -0.036           & 0.087              & 0.074              & 0.079              \\
MICE-t        & 0.2           & 0.75               & -0.143        & -0.005           & -0.002           & -0.003           & 0.201              & 0.204              & 0.195              \\
MICE-t        & 0.7           & 0.75               & 0.142         & 0                & 0.005            & 0.001            & -0.021             & -0.025             & -0.025             \\
MICE-t        & 0.7           & 0.25               & -0.065        & -0.061           & -0.065           & -0.064           & 0.103              & 0.117              & 0.106              \\
\end{tabular}
\end{table}
\end{singlespace}

\begin{singlespace}
\begin{table}[]
\centering
\caption{Median bias for $\alpha_{11}$, $\sigma^{2}_{1}$, $\sigma^{2}_{2}$, $\sigma^{2}_{3}$, $\lambda_{11}^{2}$, $\lambda_{12}^{2}$, and $\lambda_{13}^{2}$ when the missingness mechanism is ATMAR, missing percentage is 30\%, and true $\gamma$ = 0.3. MNAR missingness behaves similarly to ATMAR missingness and was, thus, excluded.}
\label{tab:my-table}
\begin{tabular}{cccccccccc}
Imputation    & True $\alpha$ & True $\lambda^{2}$ & $\alpha_{11}$ & $\sigma_{1}^{2}$ & $\sigma_{2}^{2}$ & $\sigma_{3}^{2}$ & $\lambda_{11}^{2}$ & $\lambda_{12}^{2}$ & $\lambda_{13}^{2}$ \\ \hline
MICE-tn       & 0.2           & 0.75               & -0.14         & 0.003            & 0.001            & 0                & 0.145              & 0.148              & 0.146              \\
MICE-tn       & 0.7           & 0.75               & 0.103         & 0.001            & -0.004           & -0.003           & -0.32              & -0.312             & -0.322             \\
MICE-tn       & 0.2           & 0.25               & -0.236        & -0.054           & -0.037           & -0.029           & 0.124              & 0.104              & 0.093              \\
MICE-tn       & 0.7           & 0.25               & -0.011        & -0.039           & -0.039           & -0.057           & -0.007             & -0.008             & 0.004              \\
MICE-t        & 0.2           & 0.75               & -0.144        & 0                & 0.001            & 0.001            & 0.176              & 0.147              & 0.158              \\
MICE-t        & 0.7           & 0.75               & 0.078         & 0.002            & 0.003            & -0.001           & -0.267             & -0.268             & -0.265             \\
MICE-t        & 0.7           & 0.25               & -0.041        & -0.033           & -0.049           & -0.033           & 0.009              & 0.019              & -0.001             \\
\end{tabular}
\end{table}
\end{singlespace}

\section{Coverage}

\begin{singlespace}
\begin{table}[H]
\centering
\caption{Coverage, or percentage of the true parameters which fall within the confidence interval, for $\alpha$, $\gamma$, and $\lambda^{2}$ for the $\gamma$ = 0, 30\% missingness, and TMAR mechanism condition. MCAR and MAR were excluded as they behave similarly to TMAR.}
\label{tab:my-table}
\begin{tabular}{cccccc}
Imputation    & True $\alpha$ & True $\lambda^{2}$ & $\alpha$ & $\gamma$ & $\lambda^{2}$ \\ \hline
EM-ARIMA      & 0.2           & 0.75               & 83       & 96       & 84.667        \\
EM-ARIMA      & 0.2           & 0.25               & 84.848   & 98.99    & 49.832        \\
EM-ARIMA      & 0.7           & 0.75               & 88.889   & 98.99    & 82.828        \\
EM-ARIMA      & 0.7           & 0.25               & 82       & 94       & 34.333        \\
EM-Regression & 0.2           & 0.75               & 71       & 95       & 86.333        \\
EM-Regression & 0.2           & 0.25               & 84       & 94       & 49.667        \\
EM-Regression & 0.7           & 0.75               & 0        & 80       & 37.333        \\
\end{tabular}
\end{table}
\end{singlespace}

\begin{singlespace}
\begin{table}[H]
\centering
\caption{Coverage, or percentage of the true parameters which fall within the confidence interval, for $\alpha$, $\gamma$, and $\lambda^{2}$ for the $\gamma$ = 0.3, 30\% missingness, and TMAR mechanism condition. MCAR and MAR were excluded as they behave similarly to TMAR.}
\label{tab:my-table}
\begin{tabular}{cccccc}
Imputation    & True $\alpha$ & True $\lambda^{2}$ & $\alpha$ & $\gamma$ & $\lambda^{2}$ \\ \hline
EM-ARIMA      & 0.2           & 0.75               & 81       & 83       & 89.667        \\
EM-ARIMA      & 0.2           & 0.25               & 85.859   & 91.919   & 46.801        \\
EM-ARIMA      & 0.7           & 0.75               & 88.889   & 55.556   & 64.983        \\
EM-ARIMA      & 0.7           & 0.25               & 76       & 89       & 33.333        \\
EM-Regression & 0.2           & 0.75               & 77       & 85       & 89            \\
EM-Regression & 0.2           & 0.25               & 80       & 84       & 42.667        \\
EM-Regression & 0.7           & 0.75               & 0        & 84       & 33.333        \\
\end{tabular}
\end{table}
\end{singlespace}

\begin{singlespace}
\begin{table}[H]
\centering
\caption{Coverage, or percentage of the true parameters which fall within the confidence interval, for $\alpha$, $\gamma$, and $\lambda^{2}$ for the $\gamma$ = 0, 30\% missingness, and ATMAR mechanism condition. MNAR was excluded as it behaves similarly to ATMAR.}
\label{tab:my-table}
\begin{tabular}{cccccc}
Imputation    & True $\alpha$ & True $\lambda^{2}$ & $\alpha$ & $\gamma$ & $\lambda^{2}$ \\ \hline
EM-ARIMA      & 0.2           & 0.75               & 0        & 99       & 3             \\
EM-ARIMA      & 0.2           & 0.25               & 10       & 96       & 40.667        \\
EM-ARIMA      & 0.7           & 0.75               & 62       & 96       & 4             \\
EM-ARIMA      & 0.7           & 0.25               & 31       & 98       & 74.667        \\
EM-Regression & 0.2           & 0.75               & 4        & 96       & 12.667        \\
EM-Regression & 0.2           & 0.25               & 27       & 97       & 37.333        \\
EM-Regression & 0.7           & 0.75               & 58.586   & 75       & 59.667        \\
\end{tabular}
\end{table}
\end{singlespace}

\begin{singlespace}
\begin{table}[H]
\centering
\caption{Coverage, or percentage of the true parameters which fall within the confidence interval, for $\alpha$, $\gamma$, and $\lambda^{2}$ for the $\gamma$ = 0.3, 30\% missingness, and ATMAR mechanism condition. MNAR was excluded as it behaves similarly to ATMAR.}
\label{tab:my-table}
\begin{tabular}{cccccc}
Imputation    & True $\alpha$ & True $\lambda^{2}$ & $\alpha$ & $\gamma$ & $\lambda^{2}$ \\ \hline
EM-ARIMA      & 0.2           & 0.75               & 1        & 12       & 8             \\
EM-ARIMA      & 0.2           & 0.25               & 9        & 67       & 47            \\
EM-ARIMA      & 0.7           & 0.75               & 22.449   & 1.02     & 28.231        \\
EM-ARIMA      & 0.7           & 0.25               & 7.071    & 4.04     & 67.34         \\
EM-Regression & 0.2           & 0.75               & 3        & 11       & 32.667        \\
EM-Regression & 0.2           & 0.25               & 24       & 62       & 41.667        \\
EM-Regression & 0.7           & 0.75               & 12.5     & 30.682   & 35.955        \\
\end{tabular}
\end{table}
\end{singlespace}

\section{Measurement error: $\sigma$}

\begin{figure}[H]
\centering
\includegraphics[width=\textwidth]{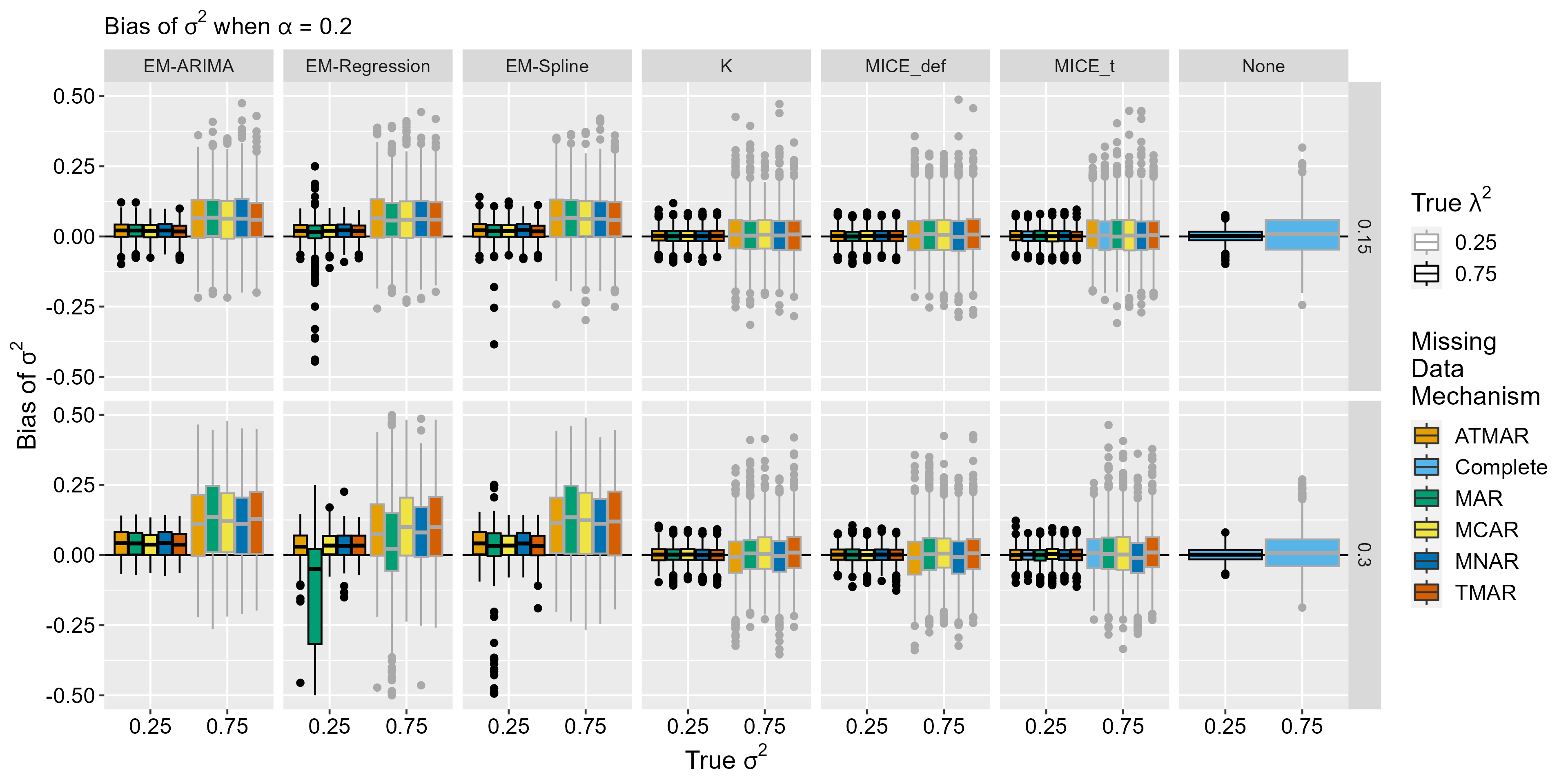}
\caption{The above graphs shows the bias of the estimated $\sigma^2$ (on the y-axis) by the true $\sigma^2$ (on the x-axis for $\alpha$ = .2. The respective missingness mechanisms are shown: complete data (orange), MCAR (sky blue), TMAR (green), MAR (yellow), MNAR (dark blue), and ATMAR (dark orange). The outlines on the box plots show the differing loading conditions: light grey for low loadings/high measurement error and black for high loadings/low measuremenent error. As the y-axis was restricted to range from -.5 to .5, 722 outliers were removed, primarily from the EM-Regression condition. For the box plots, the bottom of the box is the first quartile, the central line is the median, the top of the box is the third quartile, the whiskers extend 1.5(Interquartile Range), and the dots beyond the whiskers are outliers.}
\end{figure}

\begin{figure}[H]
\centering
\includegraphics[width=\textwidth]{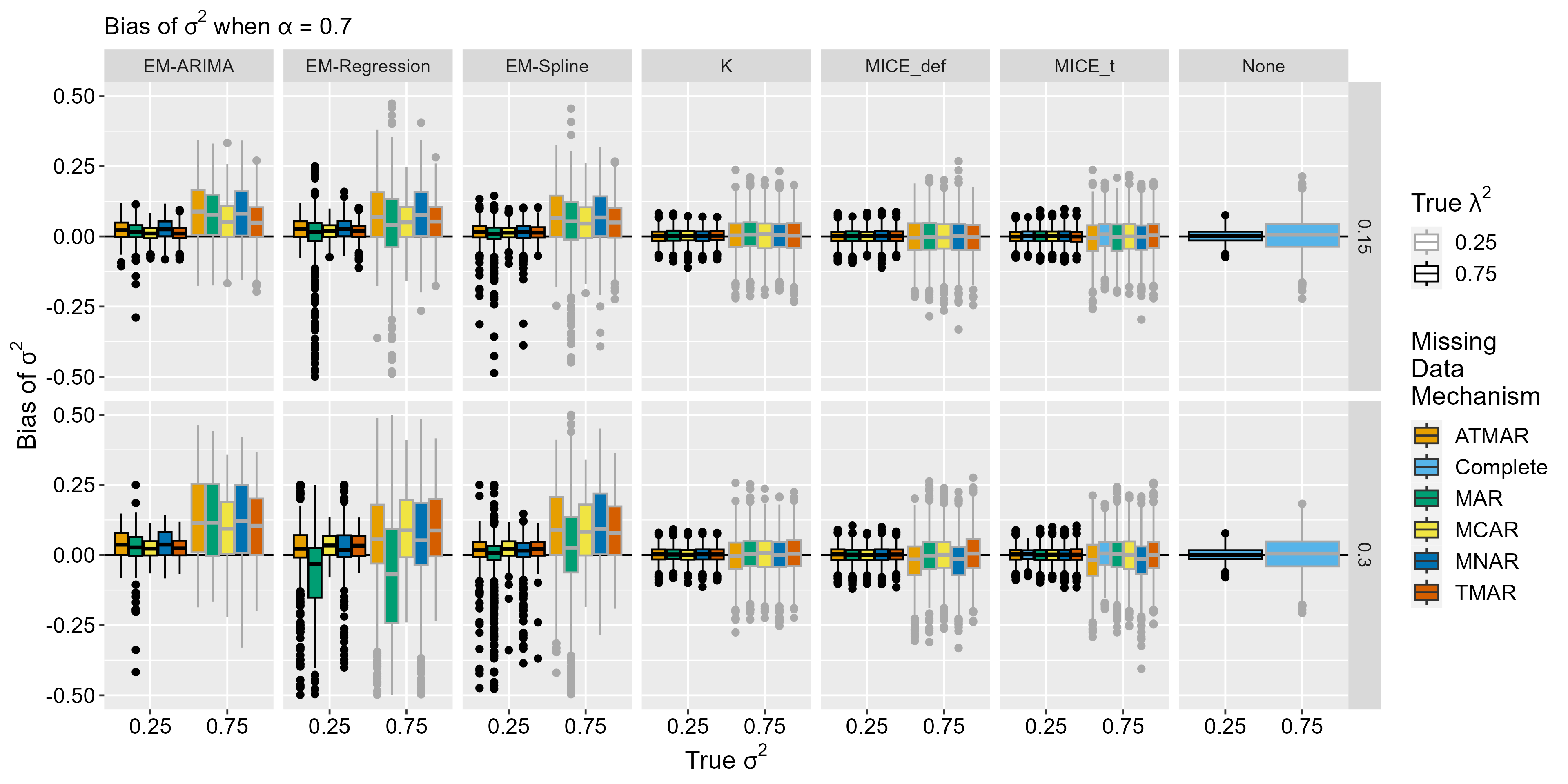}
\caption{The above graphs shows the bias of the estimated $\sigma^2$ (on the y-axis) by the true $\sigma^2$ (on the x-axis for $\alpha$ = .7. The respective missingness mechanisms are shown: complete data (orange), MCAR (sky blue), TMAR (green), MAR (yellow), MNAR (dark blue), and ATMAR (dark orange). The outlines on the box plots show the differing loading conditions: light grey for low loadings/high measurement error and black for high loadings/low measuremenent error. As the y-axis was restricted to range from -.5 to .5, 2822 outliers were removed, primarily from the EM-Regression condition. For the box plots, the bottom of the box is the first quartile, the central line is the median, the top of the box is the third quartile, the whiskers extend 1.5(Interquartile Range), and the dots beyond the whiskers are outliers.}
\end{figure}

\section{Loadings: $\lambda$}

\begin{figure}[H]
\centering
\includegraphics[width=\textwidth]{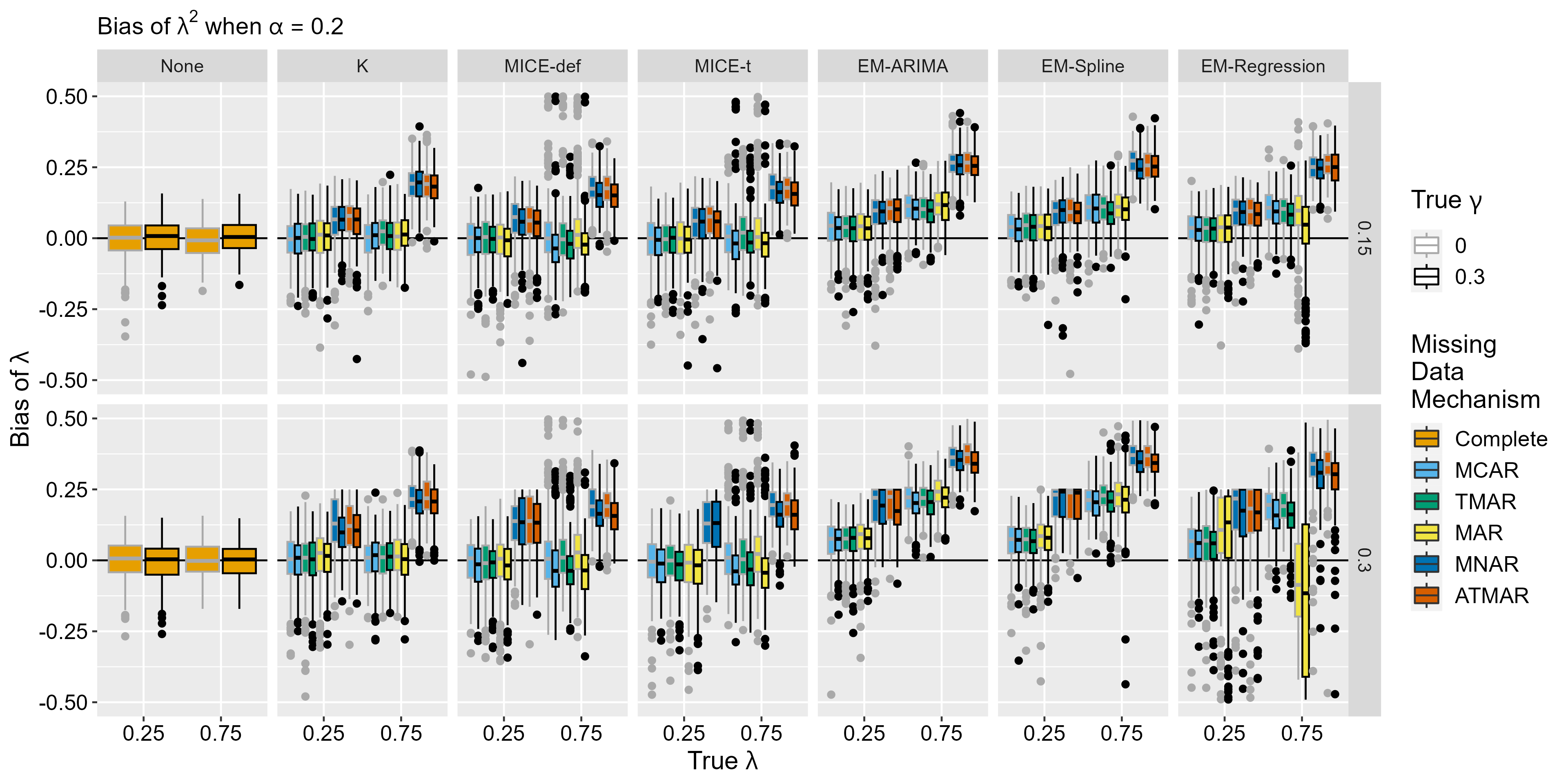}
\caption{The above graphs shows the bias of the estimated $\lambda^2$ (on the y-axis) by the true $\lambda^2$ (on the x-axis for $\alpha$ = .2. The respective missingness mechanisms are shown: complete data (orange), MCAR (sky blue), TMAR (green), MAR (yellow), MNAR (dark blue), and ATMAR (dark orange). The outlines on the box plots show the differing cross-lagged conditions: light grey for no relation and black for a strong relation. As the y-axis was restricted to range from -.5 to .5, 939 outliers were removed, primarily from the EM-Regression condition. For the box plots, the bottom of the box is the first quartile, the central line is the median, the top of the box is the third quartile, the whiskers extend 1.5(Interquartile Range), and the dots beyond the whiskers are outliers.}
\end{figure}

\begin{figure}[H]
\centering
\includegraphics[width=\textwidth]{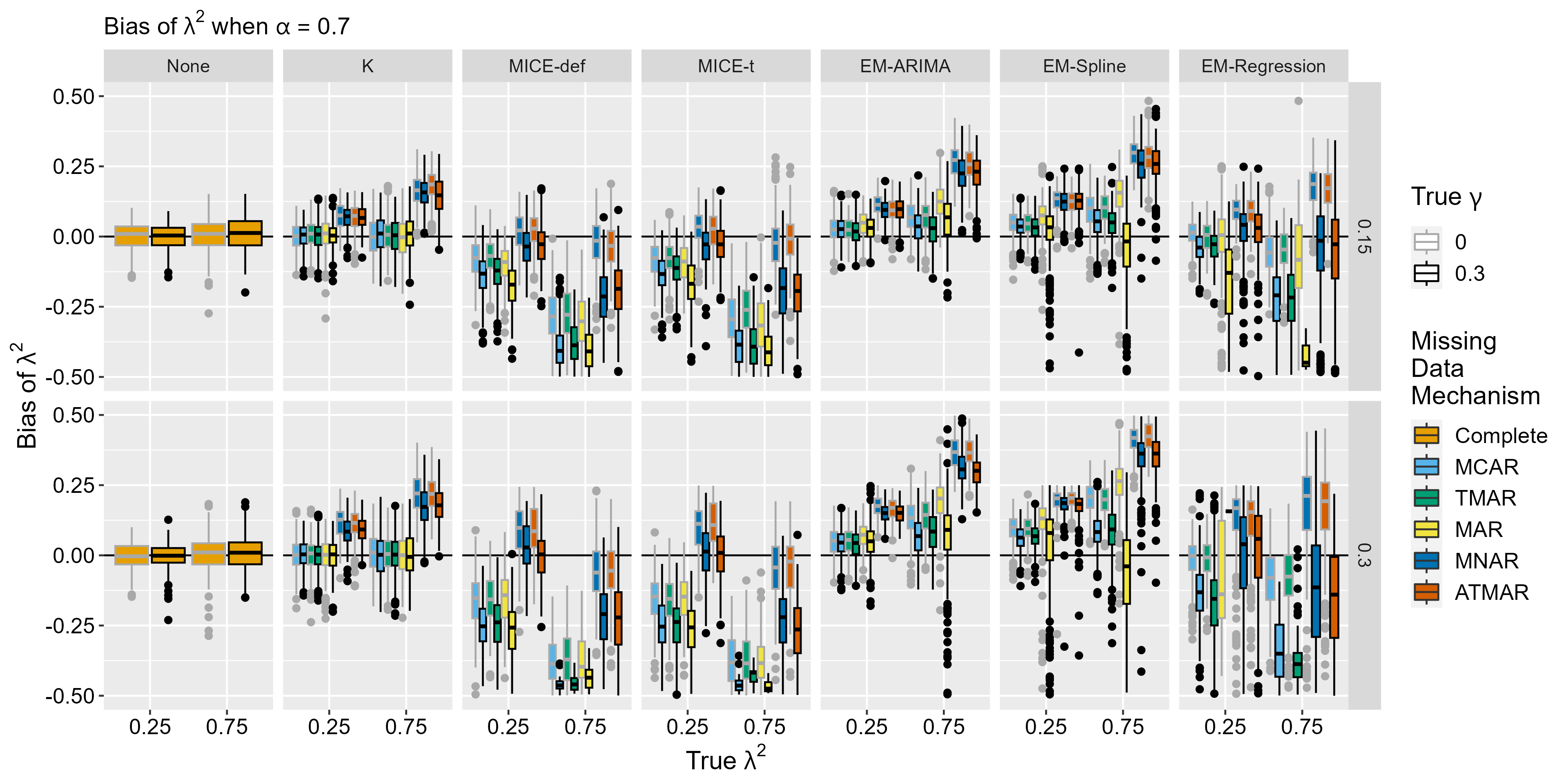}
\caption{The above graphs shows the bias of the estimated $\lambda^2$ (on the y-axis) by the true $\lambda^2$ (on the x-axis for $\alpha$ = .7. The respective missingness mechanisms are shown: complete data (orange), MCAR (sky blue), TMAR (green), MAR (yellow), MNAR (dark blue), and ATMAR (dark orange). The outlines on the box plots show the differing cross-lagged conditions: light grey for no relation and black for a strong relation. As the y-axis was restricted to range from -.5 to .5, 6790 outliers were removed, primarily from the EM-Regression condition. For the box plots, the bottom of the box is the first quartile, the central line is the median, the top of the box is the third quartile, the whiskers extend 1.5(Interquartile Range), and the dots beyond the whiskers are outliers.}
\end{figure}

\section{Standard errors}

\begin{singlespace}
\begin{table}[H]
\centering
\caption{Standard error for $\alpha_{11}$, $\sigma_{1}^{2}$, $\sigma_{2}^{2}$, $\sigma_{3}^{2}$, $\lambda_{11}^{2}$, $\lambda_{12}^{2}$, and $\lambda_{13}^{2}$ for the $\gamma$ = 0, TMAR missingness, and 30\% missingness condition. MCAR and MAR were excluded as they behave similarly to TMAR.}
\label{tab:my-table}
\begin{tabular}{cccccccccc}
Imputation    & True $\alpha$ & True $\lambda^{2}$ & $\alpha_{11}$ & $\sigma_{1}^{2}$ & $\sigma_{2}^{2}$ & $\sigma_{3}^{2}$ & $\lambda_{11}^{2}$ & $\lambda_{12}^{2}$ & $\lambda_{13}^{2}$ \\ \hline
EM-ARIMA      & 0.2           & 0.75               & 0.048         & 0.101            & 0.1              & 0.1              & 0.031              & 0.031              & 0.031              \\
EM-ARIMA      & 0.2           & 0.25               & 0.086         & 0.11             & 0.117            & 0.108            & 0.059              & 0.061              & 0.06               \\
EM-ARIMA      & 0.7           & 0.75               & 0.034         & 0.096            & 0.096            & 0.097            & 0.031              & 0.031              & 0.031              \\
EM-ARIMA      & 0.7           & 0.25               & 0.051         & 0.087            & 0.086            & 0.086            & 0.042              & 0.042              & 0.042              \\
EM-Regression & 0.2           & 0.75               & 0.049         & 0.101            & 0.102            & 0.101            & 0.032              & 0.032              & 0.032              \\
EM-Regression & 0.2           & 0.25               & 0.12          & 0.132            & 0.118            & 0.119            & 0.064              & 0.064              & 0.063              \\
EM-Regression & 0.7           & 0.75               & 0.042         & 0.099            & 0.099            & 0.097            & 0.035              & 0.035              & 0.035              \\
\end{tabular}
\end{table}
\end{singlespace}

\begin{singlespace}
\begin{table}[H]
\centering
\caption{Standard error for $\alpha_{11}$, $\sigma_{1}^{2}$, $\sigma_{2}^{2}$, $\sigma_{3}^{2}$, $\lambda_{11}^{2}$, $\lambda_{12}^{2}$, and $\lambda_{13}^{2}$ for the $\gamma$ = 0.3, TMAR missingness, and 30\% missingness condition. MCAR and MAR were excluded as they behave similarly to TMAR.}
\label{tab:my-table}
\begin{tabular}{cccccccccc}
Imputation    & True $\alpha$ & True $\lambda^{2}$ & $\alpha_{11}$ & $\sigma_{1}^{2}$ & $\sigma_{2}^{2}$ & $\sigma_{3}^{2}$ & $\lambda_{11}^{2}$ & $\lambda_{12}^{2}$ & $\lambda_{13}^{2}$ \\ \hline
EM-ARIMA      & 0.2           & 0.75               & 0.047         & 0.1              & 0.101            & 0.1              & 0.031              & 0.031              & 0.031              \\
EM-ARIMA      & 0.2           & 0.25               & 0.082         & 0.101            & 0.105            & 0.105            & 0.055              & 0.055              & 0.056              \\
EM-ARIMA      & 0.7           & 0.75               & 0.03          & 0.095            & 0.095            & 0.096            & 0.032              & 0.032              & 0.032              \\
EM-ARIMA      & 0.7           & 0.25               & 0.041         & 0.084            & 0.082            & 0.083            & 0.038              & 0.037              & 0.037              \\
EM-Regression & 0.2           & 0.75               & 0.048         & 0.1              & 0.101            & 0.1              & 0.032              & 0.032              & 0.032              \\
EM-Regression & 0.2           & 0.25               & 0.109         & 0.147            & 0.113            & 0.105            & 0.059              & 0.058              & 0.057              \\
EM-Regression & 0.7           & 0.75               & 0.044         & 0.102            & 0.098            & 0.098            & 0.045              & 0.046              & 0.045              \\
\end{tabular}
\end{table}
\end{singlespace}

\begin{singlespace}
\begin{table}[H]
\centering
\caption{Standard error for $\alpha_{11}$, $\sigma_{1}^{2}$, $\sigma_{2}^{2}$, $\sigma_{3}^{2}$, $\lambda_{11}^{2}$, $\lambda_{12}^{2}$, and $\lambda_{13}^{2}$ for the $\gamma$ = 0, ATMAR missingness, and 30\% missingness condition. MNAR was excluded as it behaves similarly to ATMAR.}
\label{tab:my-table}
\begin{tabular}{cccccccccc}
Imputation    & True $\alpha$ & True $\lambda^{2}$ & $\alpha_{11}$ & $\sigma_{1}^{2}$ & $\sigma_{2}^{2}$ & $\sigma_{3}^{2}$ & $\lambda_{11}^{2}$ & $\lambda_{12}^{2}$ & $\lambda_{13}^{2}$ \\ \hline
EM-ARIMA      & 0.2           & 0.75               & 0.046         & 0.096            & 0.098            & 0.098            & 0.028              & 0.028              & 0.028              \\
EM-ARIMA      & 0.2           & 0.25               & 0.059         & 0.078            & 0.079            & 0.078            & 0.037              & 0.037              & 0.038              \\
EM-ARIMA      & 0.7           & 0.75               & 0.032         & 0.092            & 0.094            & 0.093            & 0.027              & 0.027              & 0.027              \\
EM-ARIMA      & 0.7           & 0.25               & 0.045         & 0.077            & 0.078            & 0.078            & 0.041              & 0.042              & 0.041              \\
EM-Regression & 0.2           & 0.75               & 0.047         & 0.098            & 0.099            & 0.098            & 0.029              & 0.029              & 0.029              \\
EM-Regression & 0.2           & 0.25               & 0.066         & 3.781            & 0.083            & 0.085            & 0.041              & 0.04               & 0.041              \\
EM-Regression & 0.7           & 0.75               & 0.047         & 0.555            & 0.147            & 0.69             & 0.055              & 0.046              & 0.048              \\
\end{tabular}
\end{table}
\end{singlespace}

\begin{singlespace}
\begin{table}[H]
\centering
\caption{Standard error for $\alpha_{11}$, $\sigma_{1}^{2}$, $\sigma_{2}^{2}$, $\sigma_{3}^{2}$, $\lambda_{11}^{2}$, $\lambda_{12}^{2}$, and $\lambda_{13}^{2}$ for the $\gamma$ = 0.3, ATMAR missingness, and 30\% missingness condition. MNAR was excluded as it behaves similarly to ATMAR.}
\label{tab:my-table}
\begin{tabular}{cccccccccc}
Imputation    & True $\alpha$ & True $\lambda^{2}$ & $\alpha_{11}$ & $\sigma_{1}^{2}$ & $\sigma_{2}^{2}$ & $\sigma_{3}^{2}$ & $\lambda_{11}^{2}$ & $\lambda_{12}^{2}$ & $\lambda_{13}^{2}$ \\ \hline
EM-ARIMA      & 0.2           & 0.75               & 0.044         & 0.098            & 0.098            & 0.096            & 0.028              & 0.028              & 0.028              \\
EM-ARIMA      & 0.2           & 0.25               & 0.068         & 0.078            & 0.083            & 0.083            & 0.04               & 0.042              & 0.041              \\
EM-ARIMA      & 0.7           & 0.75               & 0.029         & 0.094            & 0.093            & 0.093            & 0.028              & 0.028              & 0.028              \\
EM-ARIMA      & 0.7           & 0.25               & 0.034         & 0.077            & 0.077            & 0.077            & 0.036              & 0.036              & 0.036              \\
EM-Regression & 0.2           & 0.75               & 0.048         & 0.1              & 0.098            & 0.099            & 0.034              & 0.035              & 0.032              \\
EM-Regression & 0.2           & 0.25               & 0.083         & 0.089            & 2.126            & 0.089            & 0.047              & 0.045              & 0.044              \\
EM-Regression & 0.7           & 0.75               & 0.088         & 0.246            & 1.305            & 0.324            & 0.18               & 0.212              & 0.19               \\
\end{tabular}
\end{table}
\end{singlespace}

\section{Median absolute relative bias}

\begin{singlespace}
\begin{table}[H]
\centering
\caption{Median absolute relative bias for $\alpha_{11}$, $\sigma_{1}^{2}$, $\sigma_{2}^{2}$, $\sigma_{3}^{2}$, $\lambda_{11}^{2}$, $\lambda_{12}^{2}$, and $\lambda_{13}^{2}$ for the $\gamma$ = 0, TMAR missingness, and 30\% missingness condition. MCAR and MAR were excluded as they behave similarly to TMAR.}
\label{tab:my-table}
\begin{tabular}{cccccccccc}
Imputation    & True $\alpha$ & True $\lambda^{2}$ & $\alpha_{11}$ & $\sigma_{1}^{2}$ & $\sigma_{2}^{2}$ & $\sigma_{3}^{2}$ & $\lambda_{11}^{2}$ & $\lambda_{12}^{2}$ & $\lambda_{13}^{2}$ \\ \hline
MICE-t        & 0.2           & 0.75               & 0.473         & 0.067            & 0.092            & 0.088            & 0.079              & 0.091              & 0.079              \\
MICE-t        & 0.7           & 0.75               & 0.481         & 0.091            & 0.085            & 0.069            & 0.606              & 0.614              & 0.614              \\
MICE-t        & 0.2           & 0.25               & 0.464         & 0.075            & 0.077            & 0.071            & 0.198              & 0.228              & 0.212              \\
MICE-t        & 0.7           & 0.25               & 0.472         & 0.08             & 0.062            & 0.074            & 0.591              & 0.607              & 0.663              \\
MICE-def      & 0.2           & 0.75               & 0.459         & 0.088            & 0.091            & 0.071            & 0.08               & 0.078              & 0.08               \\
MICE-def      & 0.7           & 0.75               & 0.476         & 0.076            & 0.082            & 0.097            & 0.578              & 0.602              & 0.606              \\
MICE-def      & 0.2           & 0.25               & 0.495         & 0.076            & 0.061            & 0.082            & 0.212              & 0.229              & 0.289              \\
\end{tabular}
\end{table}
\end{singlespace}

\begin{singlespace}
\begin{table}[H]
\centering
\caption{Median absolute relative bias for $\alpha_{11}$, $\sigma_{1}^{2}$, $\sigma_{2}^{2}$, $\sigma_{3}^{2}$, $\lambda_{11}^{2}$, $\lambda_{12}^{2}$, and $\lambda_{13}^{2}$ for the $\gamma$ = 0.3, TMAR missingness, and 30\% missingness condition. MCAR and MAR were excluded as they behave similarly to TMAR.}
\label{tab:my-table}
\begin{tabular}{cccccccccc}
Imputation    & True $\alpha$ & True $\lambda^{2}$ & $\alpha_{11}$ & $\sigma_{1}^{2}$ & $\sigma_{2}^{2}$ & $\sigma_{3}^{2}$ & $\lambda_{11}^{2}$ & $\lambda_{12}^{2}$ & $\lambda_{13}^{2}$ \\ \hline
MICE-t        & 0.2           & 0.75               & 0.482         & 0.081            & 0.077            & 0.08             & 0.083              & 0.092              & 0.089              \\
MICE-t        & 0.7           & 0.75               & 0.473         & 0.082            & 0.1              & 0.083            & 1.012              & 0.985              & 1.028              \\
MICE-t        & 0.2           & 0.25               & 0.533         & 0.074            & 0.105            & 0.079            & 0.207              & 0.197              & 0.188              \\
MICE-t        & 0.7           & 0.25               & 0.481         & 0.083            & 0.065            & 0.067            & 1.029              & 0.931              & 0.976              \\
MICE-def      & 0.2           & 0.75               & 0.512         & 0.092            & 0.073            & 0.067            & 0.077              & 0.08               & 0.079              \\
MICE-def      & 0.7           & 0.75               & 0.479         & 0.077            & 0.075            & 0.073            & 0.956              & 0.966              & 0.979              \\
MICE-def      & 0.2           & 0.25               & 0.51          & 0.071            & 0.091            & 0.085            & 0.209              & 0.199              & 0.172              \\
\end{tabular}
\end{table}
\end{singlespace}

\begin{singlespace}
\begin{table}[H]
\centering
\caption{Median absolute relative bias for $\alpha_{11}$, $\sigma_{1}^{2}$, $\sigma_{2}^{2}$, $\sigma_{3}^{2}$, $\lambda_{11}^{2}$, $\lambda_{12}^{2}$, and $\lambda_{13}^{2}$ for the $\gamma$ = 0, ATMAR missingness, and 30\% missingness condition. MNAR was excluded as it behaves similarly to ATMAR.}
\label{tab:my-table}
\begin{tabular}{cccccccccc}
Imputation    & True $\alpha$ & True $\lambda^{2}$ & $\alpha_{11}$ & $\sigma_{1}^{2}$ & $\sigma_{2}^{2}$ & $\sigma_{3}^{2}$ & $\lambda_{11}^{2}$ & $\lambda_{12}^{2}$ & $\lambda_{13}^{2}$ \\ \hline
MICE-t        & 0.2           & 0.75               & 0.716         & 0.069            & 0.093            & 0.083            & 0.268              & 0.273              & 0.26               \\
MICE-t        & 0.7           & 0.75               & 0.202         & 0.071            & 0.084            & 0.076            & 0.081              & 0.086              & 0.091              \\
MICE-t        & 0.7           & 0.25               & 0.156         & 0.11             & 0.095            & 0.103            & 0.413              & 0.469              & 0.423              \\
MICE-def      & 0.2           & 0.75               & 0.693         & 0.08             & 0.082            & 0.1              & 0.255              & 0.26               & 0.246              \\
MICE-def      & 0.7           & 0.75               & 0.199         & 0.068            & 0.094            & 0.08             & 0.114              & 0.094              & 0.113              \\
MICE-def      & 0.2           & 0.25               & 1.412         & 0.091            & 0.08             & 0.087            & 0.54               & 0.584              & 0.539              \\
MICE-def      & 0.7           & 0.25               & 0.131         & 0.101            & 0.088            & 0.092            & 0.357              & 0.376              & 0.375              \\
\end{tabular}
\end{table}
\end{singlespace}

\begin{singlespace}
\begin{table}[H]
\centering
\caption{Median absolute relative bias for $\alpha_{11}$, $\sigma_{1}^{2}$, $\sigma_{2}^{2}$, $\sigma_{3}^{2}$, $\lambda_{11}^{2}$, $\lambda_{12}^{2}$, and $\lambda_{13}^{2}$ for the $\gamma$ = 0.3, ATMAR missingness, and 30\% missingness condition. MNAR was excluded as it behaves similarly to ATMAR.}
\label{tab:my-table}
\begin{tabular}{cccccccccc}
Imputation    & True $\alpha$ & True $\lambda^{2}$ & $\alpha_{11}$ & $\sigma_{1}^{2}$ & $\sigma_{2}^{2}$ & $\sigma_{3}^{2}$ & $\lambda_{11}^{2}$ & $\lmabda_{12}^{2}$ & $\lambda_{13}^{2}$ \\ \hline
MICE-t        & 0.2           & 0.75               & 0.718         & 0.079            & 0.076            & 0.093            & 0.235              & 0.196              & 0.21               \\
MICE-t        & 0.7           & 0.75               & 0.111         & 0.062            & 0.073            & 0.091            & 0.356              & 0.358              & 0.353              \\
MICE-t        & 0.7           & 0.25               & 0.103         & 0.079            & 0.082            & 0.086            & 0.239              & 0.243              & 0.276              \\
MICE-def      & 0.2           & 0.75               & 0.757         & 0.079            & 0.071            & 0.093            & 0.21               & 0.209              & 0.195              \\
MICE-def      & 0.7           & 0.75               & 0.124         & 0.078            & 0.079            & 0.076            & 0.298              & 0.304              & 0.3                \\
MICE-def      & 0.2           & 0.25               & 1.682         & 0.086            & 0.086            & 0.094            & 0.533              & 0.529              & 0.546              \\
MICE-def      & 0.7           & 0.25               & 0.084         & 0.075            & 0.083            & 0.105            & 0.209              & 0.212              & 0.246              \\
\end{tabular}
\end{table}
\end{singlespace}